\documentclass[12pt,preprint]{aastex}

\newbox\grsign \setbox\grsign=\hbox{$>$} \newdimen\grdimen \grdimen=\ht\grsign
\newbox\simlessbox \newbox\simgreatbox
\setbox\simgreatbox=\hbox{\raise.5ex\hbox{$>$}\llap
     {\lower.5ex\hbox{$\sim$}}}\ht1=\grdimen\dp1=0pt
\setbox\simlessbox=\hbox{\raise.5ex\hbox{$<$}\llap
     {\lower.5ex\hbox{$\sim$}}}\ht2=\grdimen\dp2=0pt
\def\simgreat{\mathrel{\copy\simgreatbox}}
\def\simless{\mathrel{\copy\simlessbox}}
\newbox\simppropto
\setbox\simppropto=\hbox{\raise.5ex\hbox{$\sim$}\llap
     {\lower.5ex\hbox{$\propto$}}}\ht2=\grdimen\dp2=0pt

\slugcomment{Submitted to the Astronomical Journal}

\begin{document}


\title{Population Synthesis in the Blue III. 
The Integrated Spectrum of M67 and the Spectroscopic Age of M32}


\author{Ricardo P. Schiavon}
\affil{UCO/Lick Observatory, University of California, Santa Cruz,
CA 95064}
\email{ripisc@ucolick.org}

\author{Nelson Caldwell}
\affil{Smithsonian Astrophysical Observatory, 60 Garden Street, Cambridge,
MA 02138}
\email{caldwell@cfa.harvard.edu}

\author{James A. Rose}
\affil{Department of Physics and Astronomy, CB 3255, University of North
Carolina, Chapel Hill, NC 27599}
\email{jim@physics.unc.edu}




\begin{abstract}

We construct an integrated spectrum of the intermediate age, solar
metallicity Galactic cluster M67, from individual spectroscopic
observations of bona fide cluster members. The spectrum so obtained
is used as a template to test our stellar population synthesis (SPS)
models, in an age and metallicity regime where such models remain
largely untested. As a result, we demonstrate that our models predict a
spectroscopic age of 3.5 $\pm$ 0.5 Gyr for M67, which is the same age
we obtain from fitting isochrones to the color-magnitude diagram of
the cluster.  Full consistency is reached when using either $H\beta$,
$H\gamma$ or $H\delta$ as the age indicator. We also check if the models,
when applied to the cluster integrated spectrum, predict elemental
abundances in agreement with the known detailed abundance pattern
of the cluster. The models also pass the latter test, by predicting
the abundances of iron, magnesium, carbon and nitrogen in agreement
with detailed abundance analyses of cluster stars to within 0.1 dex.
Encouraged by the high degree of consistency of our models, we apply
them to the study of the integrated spectrum of the central 3'' of
the compact elliptical galaxy M32. The resulting luminosity-weighted
age of the galaxy ranges between 2 and 3.5 Gyr, depending on the age
indicator adopted. According to our models, the center of M32 seems to
have a super-solar iron abundance, ranging between [Fe/H] $\sim +0.1$
and +0.3, depending on the spectral index adopted. The light element
magnesium seems to be underabundant in the center of M32 relative to
iron by about $\sim$ 0.1--0.2 dex, whereas the data are consistent with
nearly solar carbon and nitrogen abundances relative to iron. We find
that single age, single metallicity stellar population models with a
solar-scaled abundance pattern cannot fit all the Balmer and metal lines
in the integrated spectrum of M32. In particular, there is a systematic
trend in the sense that bluer absorption lines indicate a younger age
and a higher metallicity. This slight inconsistency can be due either
to (unaccounted for) abundance ratio effects on blue iron and Balmer
line indices, or to a spread of the ages of the stellar populations in
M32. Current stellar population models cannot break this degeneracy at
the level of accuracy required to address this problem.

\end{abstract}




\section{Introduction}

With the commissioning of both wide field spectroscopic survey
instruments (2dF, Colless et al. 2001; SDSS, York et al. 2000)
and of deep multi-aperture spectroscopy on large aperture telescopes
(e.g. DEEP survey, Davis et al. 2003, VIRMOS-VLT Deep survey, Le F\`evre
et al. 2001), it is becoming increasingly important to reliably extract
information concerning the stellar populations in galaxies from their
integrated spectra.

Stellar population synthesis (SPS) provides the only tool to estimate the
ages and metal abundances of stellar populations in distant, unresolved
galaxies. The application of SPS models to the study of unresolved
systems, however, needs to be preceded by extensive and detailed testing
against the observations of nearby, resolved systems, for which the
key stellar population (SP) parameters, such as age, metal abundances
(i.e., metallicities and abundance ratios) and initial mass function
(IMF) are known independently.  To gain confidence that modelling of
the integrated spectra of galaxies is sufficiently reliable to extract
useful information about their luminosity-weighted mean ages and chemical
compositions, it is often proposed that integrated light observations of
star clusters should be used as a testing ground for the evolutionary
synthesis models that form the backbone of integrated light analysis.
Recently, different groups have pursued this aim, by comparing SPS model
predictions to the observations of globular clusters with a range of
SP parameters (Leonardi \& Rose 2003, Maraston et al. 2003, Beasley,
Hoyle \& Sharples 2002, Puzia et al. 2002, Schiavon et al. 2002a,b).

Given the extensive evidence in recent work that the integrated light
of early-type galaxies is often dominated by intermediate-age stellar
populations (e.g., Trager et al. 2000, Kuntschner 2000, Caldwell, Rose \&
Concannon 2003), it has become of fundamental importance that SPS models
be tested in the regime of intermediate age ($\sim$ 4 Gyr) and solar
metallicity. Moreover, this is the range of SP parameters expected
in early-type galaxies at moderately high redshifts ($z \sim 1$).
Most of the previous studies, however, have concentrated on Galactic
globular clusters, thus being limited to ages older than $\sim$ 10
Gyr. More recently Leonardi \& Rose (2003) and Beasley et al. (2002)
have extended SPS model calibration efforts to younger ages by comparing
model predictions to observations of clusters from the Large Magellanic
Cloud. In general, good consistency was found between SPS ages and
metallicities and those from the literature.  However, the quality
of cluster CMDs for age determinations, as well as the variety of
isochrones used in the literature, make the precision and homogeneity
of CMD ages for LMC clusters quite low, compared with what can be
achieved with Galactic clusters. Moreover, it is difficult to check
the metallicity scale of SP models using LMC clusters because they
lack reliable metallicity determinations based on classical abundance
analysis of high resolution spectra of stellar members. To this date,
such conditions are met only for Galactic clusters, which makes them
the ideal templates for the calibration of SP models.

In order to cure this deficiency, we constructed an integrated spectrum of
M67, an old ($\sim$ 4 Gyr) Galactic open cluster with solar metallicity.
The latter spectrum is used as a template for the calibration of the
SPS models in the regime of intermediate age and solar metallicity. Our
integrated spectrum was constructed by observing individual bona fide
cluster members and coadding their spectra weighting according to their
magnitudes and an assumed IMF.


We present a detailed comparison of our SPS models to the above
integrated spectrum of M67. We find that our model spectroscopic ages
based on all Balmer lines from $H\beta$ through $H\delta$ agree with
the CMD-based cluster age to within 0.5 Gyr.  We also verify whether the
models correctly predict elemental abundances from integrated spectra,
by comparing to the well known abundances of M67 stars, taken from the
literature. Again, our models pass the test by predicting correctly the
abundances of key elements, like magnesium, carbon and nitrogen.

Our spectrum of M67 is also compared with an integrated spectrum of
the central 3'' of M32, obtained with the same instrumental setup. We
find that the two spectra look remarkably alike, indicating that the
luminosity-weighted age and metal abundances of M32 are likely to be
very close to those of M67.

The above results encouraged us to compare our model predictions to the
integrated spectrum of another benchmark of stellar population studies,
the compact elliptical galaxy M32. The resulting spectroscopic age is
somewhere between 2 and 3.5 Gyr, depending on the Balmer line adopted
in the analysis. Also depending on the iron feature used, the mean iron
abundance of M32 results slightly above solar, or strongly super-solar.
Analysis of other features, like Mgb and CN bands, shows that magnesium
is underabundant relative to iron in the center of M32, while carbon
and nitrogen have about solar abundance ratios relative to iron.

In Section 2 we describe the selection of our sample of M67 members
and the observations. In Section 3 we describe the computation of
the integrated spectrum of M67. In Section 4, a brief description
of the models is presented, and they are compared with both the CMD
and the integrated spectrum of M67.  The analysis of the integrated
spectrum of M32 is presented in Section 5, and in Section 6 we
summarize our conclusions.

\section{Observational Data}

\subsection{Sample Selection}

\subsubsection{M67 Stars}

To obtain a representative integrated spectrum of M67 built up from
observations of individual stars in the cluster, spectra and photometry
covering the entire CMD from the tip of the giant branch through the lower
main sequence are required.  Included as well must be giant branch (GB)
``clump'' stars, i.e., the core helium burning phase, as well as blue
straggler (BS) stars.  As a starting point, we used the CMD of Montgomery,
Marschall, \& Janes (1993; hereafter MMJ) to produce a candidate list
of 103 stars covering the CMD from the tip of the giant branch down to
1.5 mag below the main sequence turnoff (MSTO, V$\sim$14), as well as
several GB clump stars and a number of BS stars.  The stars were selected
for observation strictly on the basis of their positions in the CMD.
The cutoff at the fainter MS magnitudes is imposed by the combination
of required spectral resolution and S/N ratio.  As decribed below, we
filled in the lower MS with observations of solar-abundance field dwarfs.

Naturally, there will be interlopers potentially contaminating the true
M67 cluster members.  We use four methods to filter out such interlopers.
The first is to search for proper motion non-members, based on the study
carried out by Sanders (1977).  We find 16 stars in our sample with a
proper-motion membership probability of 30\% or less.  A second criterion
is radial velocity.  In addition to our own velocities, which have a
1-$\sigma$ accuracy of $\pm11$ km/s, many of our stars are observed by
either Mathieu et al.  (1986), with a precision of better than 1 km/s,
and by Scott, Friel, \& Janes (1995), with a precision similar to ours.
We find that 13 stars have radial velocities inconsistent with membership
in M67; six of these stars overlap with the proper motion non-members.

A third way to discover non-members is if the metal-abundance of the star
is inconsistent with the near-solar composition of M67.  Specifically, the
MSTO color of M67 is similar to that of the Galactic thick disk population
(Carney, Latham \& Laird 1989; Gilmore, Wyse \& Jones 1995).  In addition,
M67 is located approximately 0.5 kpc above the Galactic plane, and thus
at the distance where one can expect significant contamination from the thick
disk MSTO.  To identify thick disk MSTO stars, we measure the Lick
Fe5270 and H$\beta$ indices (Worthey et al. 1994) for all of the stars
in our sample, then find a mean relation between Fe5270 and H$\beta$.
Stars that deviate by more than --0.4 \AA \ in Fe5270, for their 
H$\beta$, are deemed to be metal-weak, with the threshold at --0.4 \AA \
determined from the fit to H$\beta$ versus [Fe/H] given in Worthey et
al. (1994).  Altogether we find six stars in the M67 turnoff region which
are metal-poor.  Five of these stars are proper motion non-members, and
three of the stars are further confirmed as radial velocity non-members.
Given that the majority of stars in our list are indeed cluster members,
it is suggestive that the proper motion and radial velocity
non-members that we find in the turnoff region are also systematically
metal-poor.  Thus we conclude that there is significant contamination
of the observed CMD of M67 with thick disk turnoff stars.

A fourth way to detect non-members is by identifying stars with
anomalous positions in the M67 CMD based on their inferred absolute
V magnitude, assuming membership in M67, and comparing to distances inferred
from their surface gravities.  Here we make use of a pair of
spectral indices, Sr~II$\lambda 4077$/Fe~I$\lambda 4045$ and
$H\delta$/Fe~I$\lambda 4045$, that have been utilized by Rose (1984;
1985a,b) to discriminate surface gravity in stars.  Briefly, the
method relies on the gravity sensitivity of the ionization
balance between singly ionized strontium and neutral iron,
with the temperature sensitivity moderated by the strength of
$H\delta$.  Using this procedure we find five stars with anomalous
CMD positions, three of which are proper motion and/or radial
velocity non-members as well.  The five cases are as follows.  The
star MMJ6496 is positioned at V=11.26 and $(B-V)$=0.62 in the CMD, thus
is located at low surface gravity if it is indeed a member of M67,
and at an unusual location, i.e., well above the MSTO, but too red
for a BS star.  However, the surface gravity determined from the
Sr~II$\lambda 4077$/Fe~I$\lambda 4045$ index indicates that it is a
dwarf, and its Fe5270 index indicates solar composition.  Thus it
appears to be a relatively nearby ($\sim$ 200 pc away) solar
composition thin disk dwarf.  The star MMJ5336 has V=12.96,
$(B-V)$=1.11, placing it in an anomalously red region of the bottom of
the SGB.  Its spectral characteristis are all consistent with a cool
thick disk (i.e., metal-poor) giant.  The star MMJ6034 has similar
characteristics to MMJ5336.  In contrast, the stars MMJ5571 and
MMJ5688 have the problem of lying blueward in color of the MSTO.
MMJ5571 has V=12.65 and $(B-V)$=0.52, while MMJ5688 has V=12.89 and
$(B-V)$=0.45.  For both of these stars the Sr~II$\lambda
4077$/Fe~I$\lambda 4045$ index indicates low surface gravity.

We have summarized the case for the non-member stars in Table
\ref{nonmembers}, where stellar IDs (following the nomenclature of
MMJ), coordinates, magnitudes and colors are listed together with the
criteria used to decide the non-membership of each star.  In Table
\ref{members_tab} we list the stars found to be members of M67, which
were used to construct the integrated spectrum of the cluster. In
Figure \ref{targets} we display the M67 bona fide member stars in the
color-magnitude diagram.

\begin{deluxetable}{llllll}
\tablecaption{Stars which were found {\it not} to be members of
M67 \label{nonmembers}}
\tablewidth{0pt}
\tablecolumns{6}
\tablehead{
\colhead{ID (MMJ\#)} & \colhead{RA(2000)} & \colhead{Dec(2000)} &
\colhead{V} & \colhead{$(B-V)$} & \colhead{Criteria\tablenotemark{a}}}
\startdata
6474 &  8 50 21.1 & 12 21 11.5 & 10.520 & 1.230 & 2 \\
6511 &  8 51 56.0 & 11 51 26.4 & 10.600 & 0.340 & 1 \\
6508 &  8 52 10.9 & 11 31 49.3 & 10.930 & 1.150 & 1 \\
6496 &  8 51 46.0 & 11 36 18.6 & 11.260 & 0.620 & 1,2,4 \\
5877 &  8 51 30.4 & 11 48 57.6 & 12.110 & 1.006 & 2 \\
6090 &  8 51 42.6 & 11 46 36.4 & 12.368 & 0.581 & 1,2,3 \\
5124 &  8 50 41.5 & 11 44 30.0 & 12.374 & 0.746 & 1,3 \\
6293 &  8 51 58.5 & 11 46 52.7 & 12.429 & 0.618 & 1,3 \\
5451 &  8 51 07.2 & 11 53 01.6 & 12.595 & 0.637 & 3 \\
5348 &  8 51 00.8 & 11 39 37.5 & 12.620 & 0.721 & 1 \\
5571 &  8 51 15.4 & 11 47 31.3 & 12.651 & 0.517 & 4 \\
5624 &  8 51 17.9 & 11 45 54.1 & 12.730 & 0.554 & 2 \\
5239 &  8 50 51.7 & 11 48 10.1 & 12.846 & 0.501 & 1 \\
5864 &  8 51 29.4 & 11 54 13.6 & 12.868 & 0.603 & 2 \\
5688 &  8 51 20.5 & 11 46 16.2 & 12.891 & 0.448 & 4 \\
6034 &  8 51 37.9 & 11 58 49.0 & 12.911 & 1.000 & 1,4 \\
5336 &  8 50 59.7 & 11 39 22.2 & 12.964 & 1.107 & 1,2,4 \\
5222 &  8 50 50.7 & 11 35 03.0 & 12.988 & 0.540 & 1,2,3 \\
6172 &  8 51 50.6 & 11 36 33.8 & 13.419 & 0.445 & 1,2 \\
5434 &  8 51 07.4 & 11 36 53.5 & 13.422 & 0.541 & 2 \\
6430 &  8 52 13.4 & 11 46 01.1 & 13.457 & 0.514 & 2 \\
5763 &  8 51 24.1 & 11 47 09.4 & 13.477 & 0.614 & 1 \\
6467 &  8 52 16.6 & 11 42 29.8 & 13.517 & 0.536 & 1 \\
5600 &  8 51 17.6 & 11 39 36.0 & 13.562 & 0.567 & 1 \\
5257 &  8 50 53.4 & 11 40 43.5 & 13.726 & 0.609 & 1,2,3 \\
6009 &  8 51 36.3 & 11 56 50.5 & 14.001 & 0.617 & 2 \\
\enddata
\tablenotetext{a}{Reasons for determination of non-membership: 1 = proper motion;
2 = radial velocity; 3 = metallicity; 4 = position in CMD}
\end{deluxetable}

\begin{deluxetable}{lllrl}
\tablecaption{Stars which were found to be members of M67, and were
used in the construction
of the cluster integrated spectrum\label{members_tab}}
\tablewidth{0pt}
\tablecolumns{5}
\tablehead{
\colhead{ID (MMJ\#)} & \colhead{RA(1950)} & \colhead{Dec(1950)} &
\colhead{V} & \colhead{$(B-V)$} }
\startdata
 6395  &  8 49 25.57 & 11 58 05.5 & 14.036 & 0.615  \\
 5249  &  8 48 09.05 & 11 51 17.9 & 13.703 & 0.561  \\
 5583  &  8 48 32.24 & 11 55 49.8 & 13.491 & 0.630  \\
 6134  &  8 49 02.38 & 11 57 45.2 & 13.381 & 0.572  \\
 5342  &  8 48 14.97 & 12 07 52.5 & 13.265 & 0.575  \\
 5741  &  8 48 38.86 & 12 00 30.4 & 13.260 & 0.464  \\
 5795  &  8 48 41.37 & 12 03 56.3 & 13.259 & 0.619  \\
 5716  &  8 48 37.99 & 11 57 58.2 & 13.183 & 0.581  \\
 5679  &  8 48 36.28 & 11 57 09.6 & 13.145 & 0.566  \\
 5169  &  8 48 02.00 & 11 54 21.7 & 13.117 & 0.542  \\
 5855  &  8 48 45.32 & 11 56 45.3 & 12.934 & 0.917  \\
 6114  &  8 49 00.70 & 11 58 04.5 & 12.934 & 0.919  \\
 5228  &  8 48 05.79 & 12 00 28.3 & 12.931 & 0.851  \\
 6259  &  8 49 12.03 & 12 01 33.9 & 12.920 & 0.983  \\
 5969  &  8 48 50.19 & 12 01 01.7 & 12.909 & 0.534  \\
 6169  &  8 49 04.65 & 12 08 10.1 & 12.906 & 0.970  \\
 5688  &  8 48 36.52 & 11 57 33.6 & 12.891 & 0.448  \\
 6408  &  8 49 27.32 & 11 56 57.3 & 12.889 & 0.817  \\
 5318  &  8 48 13.96 & 12 03 38.2 & 12.862 & 0.941  \\
 5284  &  8 48 10.12 & 12 07 44.4 & 12.844 & 0.522  \\
 5756  &  8 48 39.74 & 11 58 33.4 & 12.835 & 0.783  \\
 5586  &  8 48 31.59 & 12 04 15.7 & 12.831 & 0.567  \\
 5996  &  8 48 51.19 & 12 09 14.6 & 12.826 & 0.775  \\
 5118  &  8 47 56.63 & 11 59 00.8 & 12.818 & 0.521  \\
 5248  &  8 48 07.54 & 12 08 11.0 & 12.803 & 0.547  \\
 6313  &  8 49 15.49 & 12 04 16.9 & 12.790 & 0.584  \\
 5833  &  8 48 44.05 & 12 00 45.0 & 12.784 & 0.487  \\
 5350  &  8 48 15.96 & 12 05 48.0 & 12.782 & 0.813  \\
 5927  &  8 48 48.33 & 11 59 19.0 & 12.777 & 0.822  \\
 6107  &  8 48 59.97 & 11 57 42.9 & 12.750 & 0.758  \\
 5362  &  8 48 16.91 & 12 01 27.1 & 12.725 & 0.739  \\
 5993  &  8 48 51.97 & 11 57 51.6 & 12.722 & 0.683  \\
 5059  &  8 47 52.01 & 11 54 30.2 & 12.712 & 0.913  \\
 5191  &  8 48 03.55 & 11 56 05.6 & 12.700 & 0.482  \\
 6228  &  8 49 08.23 & 12 15 41.8 & 12.688 & 0.622  \\
 5853  &  8 48 44.67 & 12 03 17.6 & 12.677 & 0.684  \\
 5571  &  8 48 31.36 & 11 58 48.4 & 12.651 & 0.517  \\
 5929  &  8 48 48.38 & 11 59 10.2 & 12.648 & 0.617  \\
 5041  &  8 47 50.31 & 11 51 11.4 & 12.629 & 0.594  \\
 5643  &  8 48 34.61 & 11 58 19.8 & 12.596 & 0.783  \\
 5451  &  8 48 23.04 & 12 04 18.3 & 12.595 & 0.637  \\
 5790  &  8 48 41.30 & 11 58 51.7 & 12.533 & 0.589  \\
 6089  &  8 48 58.36 & 12 01 10.3 & 12.514 & 0.598  \\
 6158  &  8 49 04.22 & 12 02 30.6 & 12.505 & 0.663  \\
 5544  &  8 48 29.47 & 12 01 54.9 & 12.286 & 0.676  \\
 5699  &  8 48 37.17 & 11 57 09.9 & 12.260 & 0.569  \\
 5997  &  8 48 51.62 & 12 04 52.9 & 12.230 & 0.993  \\
 5667  &  8 48 35.84 & 11 58 17.5 & 12.126 & 0.458 \\
 6477  &  8 48 11.49 & 12 03 30.3 & 12.04  & 0.60   \\
 6502  &  8 48 58.23 & 12 01 26.0 & 11.63  & 1.05   \\
 6484  &  8 48 23.70 & 11 59 25.7 & 11.55  & 0.41   \\
 6488  &  8 48 39.66 & 12 01 06.6 & 11.52  & 0.87   \\
 6505  &  8 48 58.23 & 12 02 41.3 & 11.33  & 1.07   \\
 6491  &  8 48 37.60 & 12 03 55.1 & 11.315 & 0.61   \\
 6489  &  8 48 32.90 & 12 02 03.4 & 11.20  & 1.08   \\
 6480  &  8 48 30.30 & 11 56 17.3 & 11.078 & 0.43   \\
 6501  &  8 48 48.51 & 12 00 09.9 & 11.063 & 0.19   \\
 6490  &  8 48 42.88 & 12 03 10.1 & 10.99  & 0.11   \\
 6497  &  8 48 59.52 & 11 55 44.9 & 10.76  & 1.13   \\
 6510  &  8 49 26.73 & 11 55 25.2 & 10.70  & 0.11   \\
 6492  &  8 48 28.53 & 12 03 59.1 & 10.59  & 1.12  \\
 6506  &  8 48 59.68 & 12 08 00.8 & 10.58  & 1.10   \\
 6512  &  8 49 15.35 & 12 06 24.0 & 10.55  & 1.10   \\
 6503  &  8 48 44.87 & 12 01 50.7 & 10.55  & 1.12   \\
 6494  &  8 48 42.01 & 12 05 09.4 & 10.48  & 1.10   \\
 6485  &  8 48 38.71 & 11 59 19.0 & 10.48  & 1.11   \\
 6516  &  8 49 34.57 & 11 55 46.7 & 10.47  & 1.12   \\
 6481  &  8 48 27.72 & 11 56 38.8 & 10.03  &-0.073  \\
 6472  &  8 47 34.04 & 12 06 34.9 & 9.98   &1.10   \\
\enddata
\end{deluxetable}

\subsubsection{Lower Main Sequence Field Stars}

With the instrumental setup available (see Section \ref{spectra}) we
cannot easily go fainter than V=14, so we need to fill in the lower main
sequence by observing field stars of roughly solar abundance.  Since these
later-type stars all have lifetimes in excess of the age of the Galaxy,
they are consequently unevolved.  Also, they will be seen to contribute
only modestly to the integrated spectral indices in the blue.  Thus we
do not consider that using field stars as surrogates for M67 members in
the K dwarf regime produces any serious concerns. The stars adopted
re listed in Table \ref{field_stars}, where the V magnitudes listed
are not the stars' apparent magnitudes but rather the magnitudes they
would have at the distance of M67. The latter were computed using
the Hipparcos parallaxes of the field stars and the distance modulus of
the cluster, discussed in Section \ref{cmdage}.

\subsubsection{Field M Giants}

Due to the decreasing lifetime as a star ascends the GB,
representation of the upper GB of M67 is sparse.  To remedy this
situation, we have observed a number of cool field M giants, which
were selected among stars from the Jones (1999) spectral library, on
the basis of the strength of TiO bands in their spectra. The stars
adopted are listed in Table \ref{field_stars}.

\begin{deluxetable}{lrlc}
\tablecaption{Field Stars Included in the Construction of the Integrated
Spectrum of M67. The V magnitudes listed are the apparent magnitudes these
stars would have if they were at the distance of M67 \label{field_stars}}
\tablewidth{0pt}
\tablecolumns{4}
\tablehead{
\colhead{ID} & \colhead{V} & \colhead{$(B-V)$} & \colhead{Spectral Type}}
\startdata
HD61606A &  16.055 & 1.00 & K2V  \\
HD136834 &  15.845 & 1.03 & K3V  \\
HD37216  &  15.175 & 0.78 & G5  \\
HD165341 &  15.105 & 0.90 & K0V  \\
HD69830  &  15.075 & 0.84 & K0V  \\
HD106156 &  15.055 & 0.83 & G8V  \\
HD128987 &  14.975 & 0.72 & G6V  \\
HD126511 &  14.865 & 0.94 & G5  \\
HD69582  &  14.675 & 0.74 & G5  \\
HD49178  &  14.495 & 0.70 & G0  \\
HD158614 &  13.835 & 0.76 & G9IV-V  \\
HD94705  &   9.445 & 1.25 & M5.5III  \\
HD62721  &   9.195 & 1.53 & K4III  \\
HD47914  &   9.135 & 1.54 & K5III  \\
HD70272  &   9.135 & 1.61 & K4.5III  \\
HD60522  &   9.125 & 1.57 & M0III  \\
HD98991  &  10.685 & 0.45 & F3IV  \\
\enddata
\end{deluxetable}

\subsection{Observations}  \label{spectra}

Stars in M67 were observed from 2002 January to April, using the
FAST spectrograph (Fabricant et al 1998) on the 1.5m telescope of
the Whipple Observatory. A 600 gpm grating was used with a 2" slit,
which gave a spectral coverage from 3500 ${\rm\AA}$ to 5500 ${\rm\AA}$,
at a resolution of 2.7 ${\rm\AA}$ and a dispersion of 0.75 ${\rm\AA}$
pixel$^{-1}$.  All observations were taken with the slit aligned with
the parallactic angle.  Exposure times ranged from a few seconds for
the bright giants, to 30 minutes for main sequence stars.  Spectra were
extracted from the CCD images in the standard way using programs in IRAF,
and flux calibrated using standard star spectra collected during the
observing run.  Non-photometric conditions and the relatively small slit
meant that zero point of the flux calibration is not accurate, though
the relative fluxes between wavelengths should be good to about 5\%,
based on the fits to all the spectrophotometric standard stars.

Bright field stars were observed with the same equipment and
technique, to supply a library for the fainter main sequence stars,
as well as M giants.  

We also obtained a spectrum of M32 using the same instrumental setup as
above, but adopting a 3" slit width. The integrated spectrum was obtained
through standard procedures, adopting an extraction window of 3" centered
on the peak of the galaxy's light profile along the slit.

\begin{figure}
\plotone{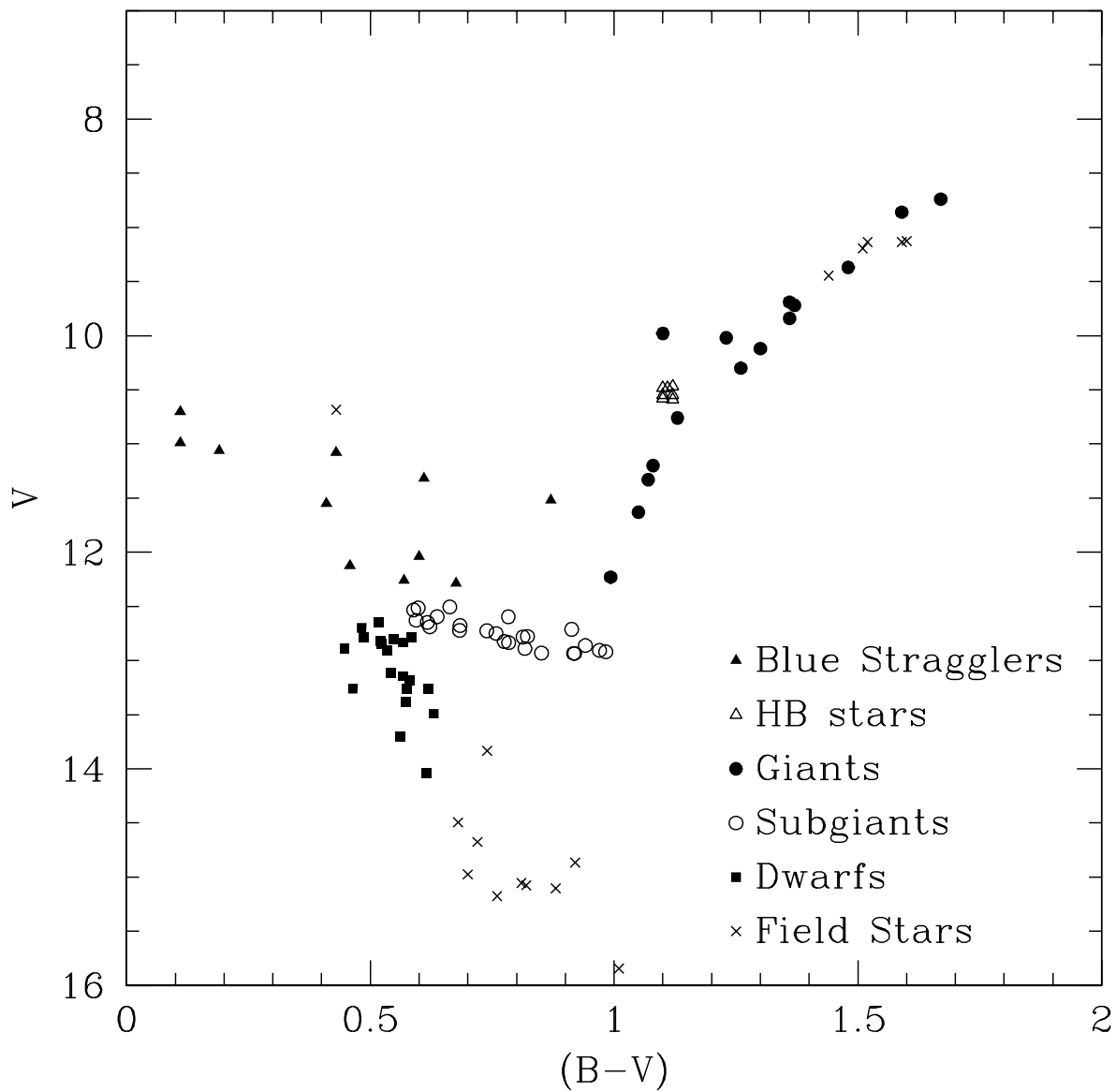}
\caption{Color-magnitude diagram showing the M67 members and field
stars observed in this work. The photometry of M67 stars was taken
from Montgomery et al. (1993).  Our observations of M67 members cover
all the evolutionary stages of relevance for integrated light studies
in the optical. Different evolutionary stages are denoted by different
symbols, indicated in the lower right panel. For completeness, we also
observed field stars in areas where the color-magnitude diagram of M67
is scarcely populated (M giants) or where the stars are too faint to be
observed efficiently (K dwarfs).  Photometry of field stars was taken
from the SIMBAD database, together with their Hipparcos parallaxes. 
The latter data were used to estimate the stars' positions in the CMD
as if they were at the distance of M67.
} \label{targets} 
\end{figure}

\subsubsection{Conversion to the Lick/IDS System} \label{conver}

\begin{figure}
\plotone{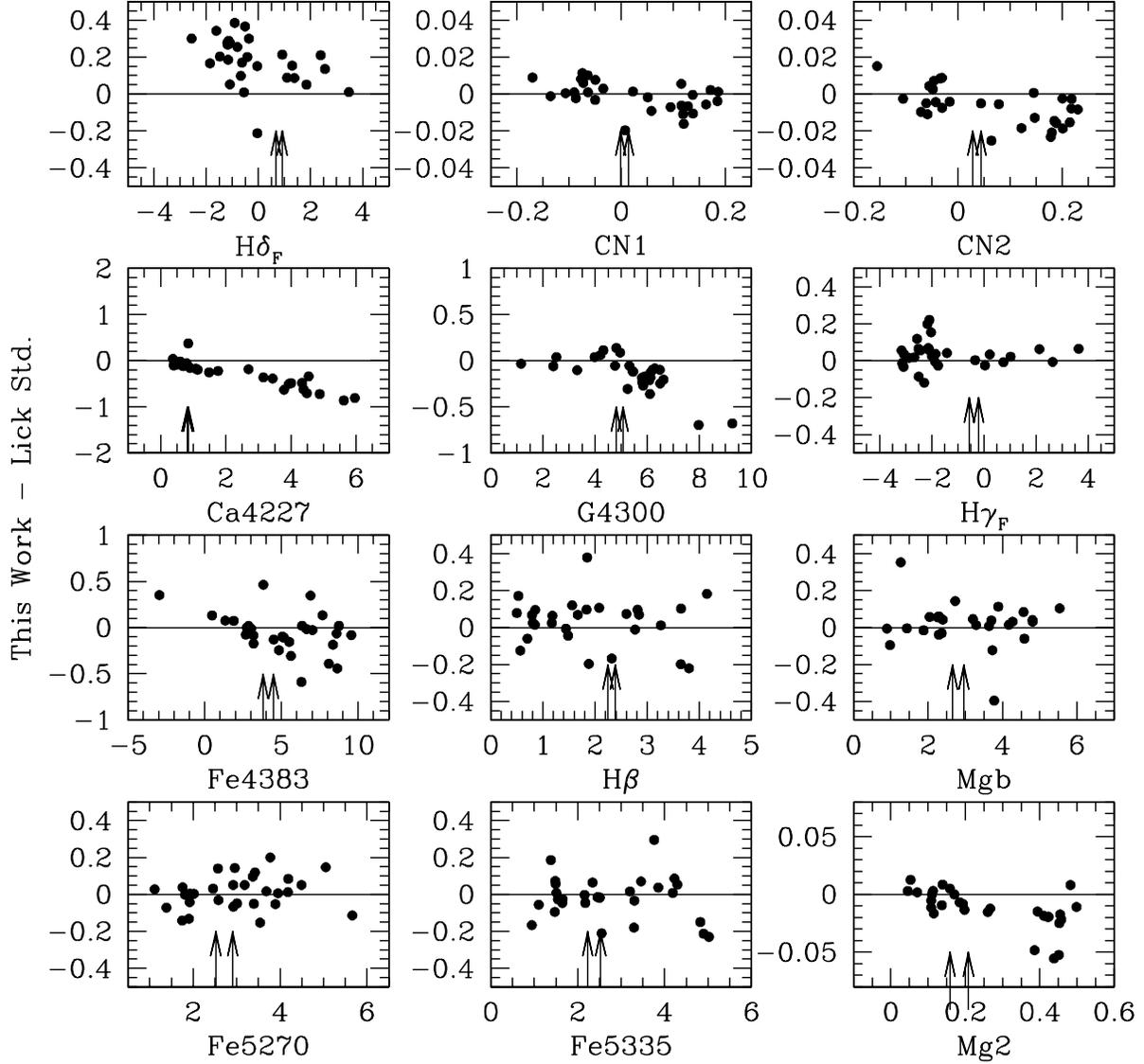}
\caption{Comparison of the line indices measured in our spectra for
Lick/IDS standards with the standard values. For all indices (except
Ca4227) only minor zeropoint corrections are needed to convert our
data to the Lick/IDS system. The arrows indicate the values measured in
the integrated spectra of M67 and M32, the main targets of this study,
showing that they are in a locus in the index space which is very well
populated by standard star measurements. The errobars of our index
measurements are about the same size as the symbols.
} 
\label{calib}
\end{figure}

In order to compare measurements of the equivalent widths (EWs) of
absorption lines in our observed spectra with model predictions, we
need to convert our measurements into the Lick/IDS system. The standard
recipe for this calibration has been described in detail by Worthey \&
Ottaviani (1997). It consists of collecting spectra of Lick/IDS standards
(Worthey et al. 1994), smoothing those spectra to the Lick/IDS resolution
(given by Worthey \& Ottaviani 1997), and comparing the EWs measured in
the smoothed spectra with the standard values.  In Figure \ref{calib}
we plot the difference between the EWs of Lick indices measured in our
standard star spectra and those measured in the spectra of the same stars
in the Jones (1999) spectral library, converted to the Lick/IDS system
(for details, see Schiavon 2003, in preparation).

Our measurements are generally in good agreement with the standard values
for most of the indices.  For some indices we find a systematic trend
of the residuals as a function of index strength. The most obvious case
is that of Ca4227, but slight trends can also be seen for $H\delta_F$,
G4300, Fe4383 and $Mg2$. The errors in our index values were estimated
from measurements made on 20 repeat spectra taken for 7 Lick standards. In
all cases the errorbars are about the same size as the symbols in Figure
\ref{calib}, indicating that most of the scatter in those plots comes
from converting between equivalent-width systems defined by the different
instrumental setups. We will henceforth base our errorbars on the {\it
r.m.s.} scatter between our data for Lick standards and the standard
values (Figure \ref{calib}).

A key aspect of such calibrations is that the calibrating standards
must encompass the range of EWs characteristic of the target
systems, otherwise systematic effects such as that seen in Figure
\ref{calib} cannot be adequately accounted for, possibly resulting
in large errors.  For this reason, we indicate by arrows in Figure
\ref{calib} the index values measured in the spectra of the two
targets of our analysis, M67 and M32.

For the indices not exhibiting any slope as a function of index
strength, only minor zero point corrections were sufficient to bring
our measurements to the Lick/IDS system. That was the case for CN1,
CN2, $H\gamma_F$, $H\beta$, Mg $b$, $Fe5270$, $Fe5335$ and $Mg2$.
For the remaining indices, the zeropoint corrections were computed as
the average value in a $\sim$ 2 ${\rm\AA}$-wide window around the values
measured for M67 and M32 (0.1 mag in the case of CN1, CN2 and $Mg2$).
On average, 15 stars were used in computing the zeropoint corrections.
We note that the index values for M67 and M32 are in regions of the
index space that are very well populated by standard stars, so that we
consider our calibration onto the Lick/IDS system to be robust. 

It is important to emphasize how accurately such calibration needs
to be done.  For instance, a $\pm 0.1 {\rm\AA}$ shift in $H\delta_F$
implies a change in the spectroscopic age of M67 by about $\mp$ 1
Gyr. $H\beta$ and $H\gamma$ are slightly less affected, but the case of
the G band (G4300) is especially delicate. There is a sharp decline of
the residuals for EWs larger than $\sim 5 {\rm\AA}$, and as a result,
if we had adopted a zeropoint correction considering all the calibrating
standards, the G4300-based spectroscopic age of M67 would be 2 Gyr older
(see discussion in Section \ref{agesect}).

In Appendix A we provide a table with Lick/IDS index measurements taken
in the spectra of all the field and M67 stars observed in this work.

\section{The Integrated Spectrum of M67} \label{comput}


In this section we describe the construction of the integrated spectrum
of M67 from the individual spectra of the stellar members and field stars
displayed in Figure \ref{targets}. The reader who is not interested in
technicalities, and would rather focus on the analysis of the integrated
spectrum is advised to skip to Section \ref{result}.

The integrated spectrum of the cluster is given by the following
expression

\begin{equation} \label{eq1}
  F(\lambda) = \sum_{i=1}^N f_i(\lambda) \times n_i \times
10^{-0.4 (B_i - B'_i)}
\end{equation}

\noindent where $f_i(\lambda)$ is the flux-calibrated spectrum of the
{\it i-th} star and $B_i$ is its $B$ magnitude taken from the Montgomery et
al. (1993) photometry. The magnitudes were corrected for interstellar
reddening assuming $E(B-V)$ = 0.05. $B'_i$ is a magnitude measured in
the observed spectrum by integrating it with a $B$-filter transmittance
curve taken from Lejeune, Cuisinier \& Buser (1997). The latter step
was taken in order to normalize all the spectra in a similar fashion,
before weighting them according to the stars' dereddened magnitudes.

The quantity $n_i$ is proportional to the number of stars at each
evolutionary stage. For all the stars from the main sequence to the
tip of the first-ascent giant branch, $n_i$ was computed assuming a
power-law IMF, normalized such that the total mass of the system is
equal to 1 $M_\odot$.  For horizontal branch stars and blue stragglers,
we followed a different procedure as explained a few paragraphs below.

In order to compute $n_i$, we need to estimate the stellar masses.
This was achieved by interpolating stellar colors or magnitudes in a
theoretical isochrone for solar metallicity and an age of 4 Gyr. The
isochrone was truncated at the tip of the first-ascent giant branch. This
procedure was followed for all stars except the M giants ($T_{eff}$ $< 4000
$K). The stars are then sorted according to their masses so that the
weight for each star was given by

\begin{equation} \label{eq2}
n_i = \int_{M_i - (\Delta M_{i})/2}^{M_i + (\Delta M_{i+1})/2} 
M^{-(1+x)} dM
\end{equation}

\noindent where $M_i$ is the mass of the $i$-th star, $x$ is the IMF
exponent, and $\Delta M_{i} = M_i - M_{i-1}$, so that the mass bin
occupied by the $i$-th star is given by

\begin{equation}   \label{eq3}
\Delta M = (M_{i+1} - M_{i-1})/2
\end{equation}

We adopt a Salpeter IMF, with $x=1.35$. It can be seen that
the weight is higher for lower mass stars, and higher for larger
$\Delta M$. The latter means that an individual star in a more densely
populated region of the CMD of Figure \ref{targets} has a lower weight
in the final spectrum because the mass bin occupied by it is narrower.
Along the giant branch, the mass varies by just a tiny fraction so that
ultimately the weight for each star is mostly dependent on $\Delta M$.
This is important for the reasons discussed in Section \ref{contm}

For blue stragglers and horizontal-branch stars, $n_i$ was computed from
the observed numbers relative to main sequence stars in the range $14.5
< m_{3890} < 15.5$ as given by the color-magnitude diagrams of Fan et
al. (1996) and Deng et al. (1999). This magnitude interval translates
into $13.649 < V < 14.427$, according to the color transformations given
by equations (3) and (4) of Fan et al.

Fan et al. (1996), analysing photometric data complete down to many
magnitudes below the turnoff, found that the present-day mass function of
M67 has an exponent $x \sim 0.1$ for masses between 0.8 and 1.2 $M_\odot$,
and that it levels off for lower masses. They interpret this result as
evidence that the cluster has undergone evaporation of low-mass stars.
We tested our computations for the effect of the power-law exponent and
found that both the overall spectral shape and the absorption line EWs
remain virtually constant even for wide variations of $x$.  Therefore,
throughout this paper we adopt a single-exponent Salpeter IMF because
our interest is in using M67 as a template single stellar population,
free of the effects of dynamical evolution.

The assumed value for the reddening towards the cluster has a small impact
in the computed integrated spectrum. This effect has been discussed
by Schiavon et al. (2002a). It is due to the fact that a variation in
$E(B-V)$ implies a change in the inferred absolute magnitudes of the
stars, which in turn affect the computed stellar masses, and thus the
relative weights given to each star in equation (\ref{eq1}). The effect
is small, and is indicated in Figures \ref{ind1} and \ref{ind2} in the
form of arrows.

\subsection{Contribution by M Giants to the Integrated Light} 
\label{contm}

For the M giants, the optical colors cannot be used for mass estimation
for two reasons: M giants are strongly variable, and optical colors
saturate for temperatures below 4000K. The strong variability implies
that a given M giant spectrum might not correspond to a measured color
if they have not been collected at approximately the same time. On the
other hand, saturation makes it difficult to use optical colors to get
masses through interpolation in the isochrones. A good example is given
by the pair of field stars HD94705 (M5.5III) and HD62721 (K4III). Their
$(B-V)$s are 1.46 and 1.53 respectively, but the effective temperatures
inferred from the strength of the TiO bands in their spectra are $\sim$
3400 and 3900 K respectively.  If their colors were used to estimate
their masses, and thence their weighting factors in equation (\ref{eq1}),
a large error would result in the case of HD94705, because being cooler
it represents a later, less populated evolutionary stage than HD62721.

Our procedure was to estimate the $T_{eff}$s of all stars redder than
$(B-V) = 1.3$ from the strength of their TiO bands, using an empirical
calibration derived from measurements in the spectra of Lick standards
whose $T_{eff}$s were determined from angular diameter measurements or
the infrared flux method (Schiavon 2003, in preparation). The $T_{eff}$s
so derived are listed in Table \ref{tbl-tef}. They were used to estimate
stellar masses and colors for stars cooler than $\sim$ 4000K, through
interpolation in the theoretical isochrone for 3.5 Gyr and [Fe/H]=0.0. The
weights of the M giants resulting from this procedure were substantially
higher than when the masses were interpolated from their $(B-V)$s, because
there are only four M giants in our sample, and therefore each of these
stars occupies a wide mass bin (equations \ref{eq2} and \ref{eq3}). As a
result, when the M giants are accounted in this way, the strength of the
redder indices, especially Mg $b$, becomes considerably stronger. In fact,
the latter index becomes $\sim$ 0.3 ${\rm\AA}$ stronger than when the
masses of M giants were (wrongly) estimated from their colors. $H\beta$
is also affected, but to a lesser degree.

\begin{deluxetable}{ccccc}
\tablecaption{Cool giants included in the synthesis of the integrated spectrum
of M67 and $T_{\rm eff}$s used to estimate their masses.
\label{tbl-tef}}
\tablewidth{0pt}
\tablecolumns{4}
\tablehead{
\colhead{ID} & $T_{\rm eff}$ (K) & & ID &$T_{\rm eff}$ (K)  }
\startdata
HD 94705 &  3350 & & MMJ-6513 & 3950 \\
MMJ-6514 &  3740 & & MMJ-6515 & 3950 \\
HD 60522 &  3900 & & MMJ-6486 & 3950 \\
MMJ-6471 &  3920 & & MMJ-6470 & 3960 \\
HD 70272 &  3930 & & MMJ-6499 & 3960 \\
HD 47914 &  3940 & & MMJ-6482 & 3960 \\
HD 62721 &  3940 & & MMJ-6495 & 3960 \\
\enddata
\end{deluxetable}

\begin{figure}
\plotone{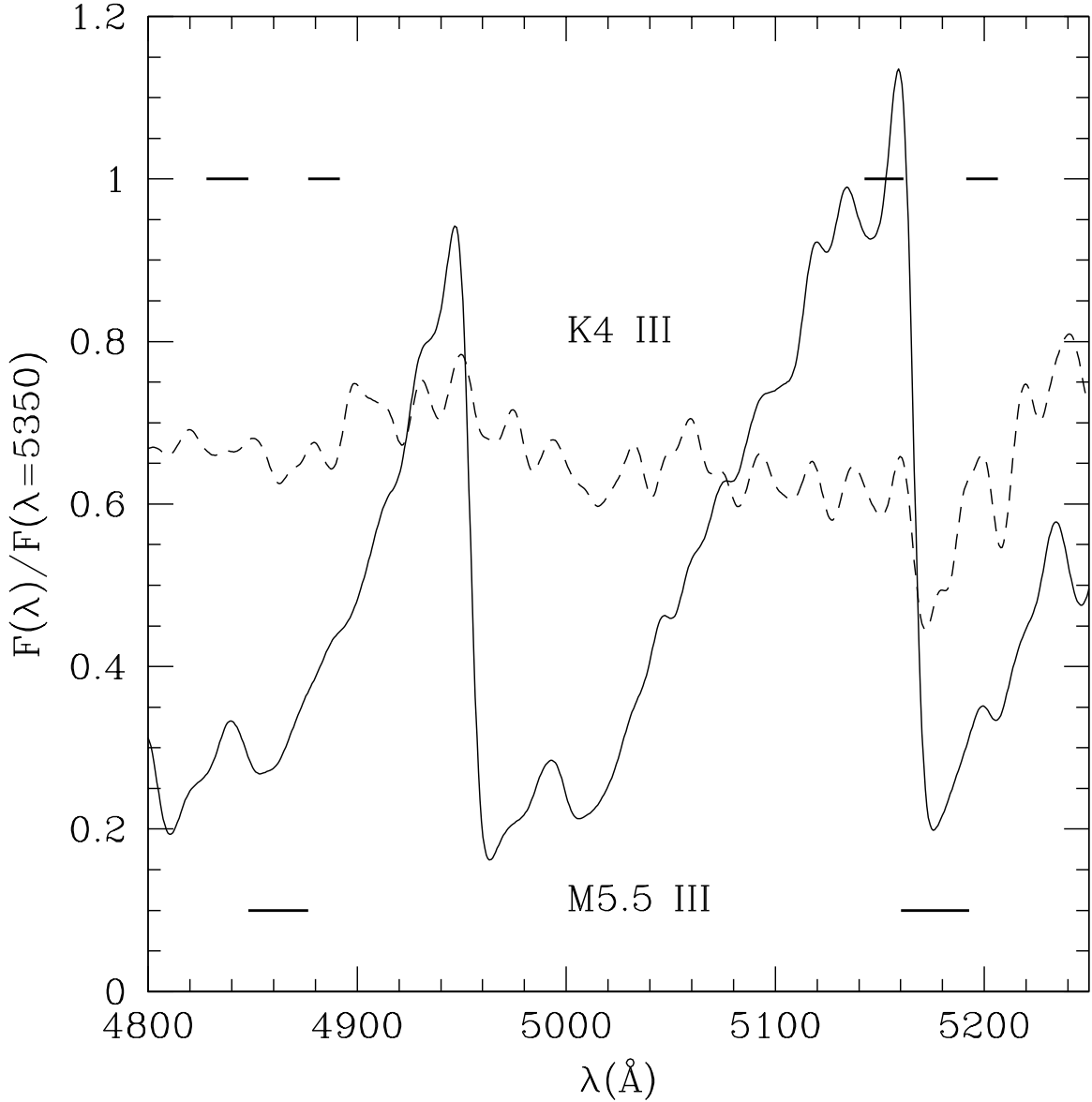}
\caption{Comparison between the spectra of two giants: HD94705 (M5.5III,
solid line) and HD62721 (K4III, dashed line), with essentially identical
$(B-V)$s. The pseudocontinua and passband of $H\beta$ (left) and Mg $b$
(right) are illustrated by the horizontal lines. The spectrum of the M
giant is characterized by very strong TiO bands, whose bandheads coincide
with the passbands and continuum windows of the Lick indices. The case
of Mg $b$ is particularly sensitive, since M giant stars contribute
roughly 6 times more flux in the blue pseudocontinuum than in the index
passband. For this reason, errors in estimating the contribution of very
cool stars to the integrated light of M67 can cause sizeable errors in
the integrated value of Mg $b$. The same comments apply, to a lesser
extent, to $H\beta$.}
\label{mgiant}
\end{figure}

The dependence of Mg $b$ and $H\beta$ on the contribution by M giants
is due to the very strong TiO bands present in their spectra, which can
sizeably affect the redder Lick indices. To illustrate this, we compare
in Figure \ref{mgiant} the spectra of HD62721 and HD94705, overlaid by
horizontal lines indicating the passband and pseudocontinuum definitions
of $H\beta$ and Mg $b$. As can be seen, strong TiO bandheads coincide
with the index passbands. In particular, in the case of Mg $b$, the M
giant spectrum is roughly 6 times brighter in the blue pseudocontinuum
window than in the index passband. Therefore, increasing the contribution
by M giants to the integrated light causes a substantial increase in
Mg $b$. As a consequence, it is very important to compute the weights
of the M giants precisely, because even though their contribution
to the total integrated light in the optical is not very large, their
influence on key spectral indices like Mg $b$ and $H\beta$ is by no means
negligible. It goes without saying that SP model predictions for such
indices are likewise afflicted by such uncertainties.  This is an issue
that is seldom appreciated in the literature, in particular by stellar
population modellers.  The situation is of course even more serious for
indices located towards the far red where the contribution of M giants to
the integrated light is dominant (e.g. Schiavon, Barbuy \& Bruzual 2000,
Schiavon \& Barbuy 1999).

Another important issue regards the luminosity function in the upper
giant branch. In particular, our procedure to compute the integrated
spectrum of M67 needs to account for stars in the asymptotic giant
branch (AGB). Recall that when we computed the weights for the stars in
equation (\ref{eq1}), we truncated the isochrone used in the computation
of stellar masses at the tip of the first-ascent giant branch. In order
to account for the AGB stars, we increase $n_i$ for stars brighter than
the horizontal branch by 0.1 dex, according to the theoretical LF of
the Padova isochrones.

\subsection{Resulting Integrated Spectrum and the Effect of Blue
Stragglers} \label{result}

\begin{figure} 
\plotone{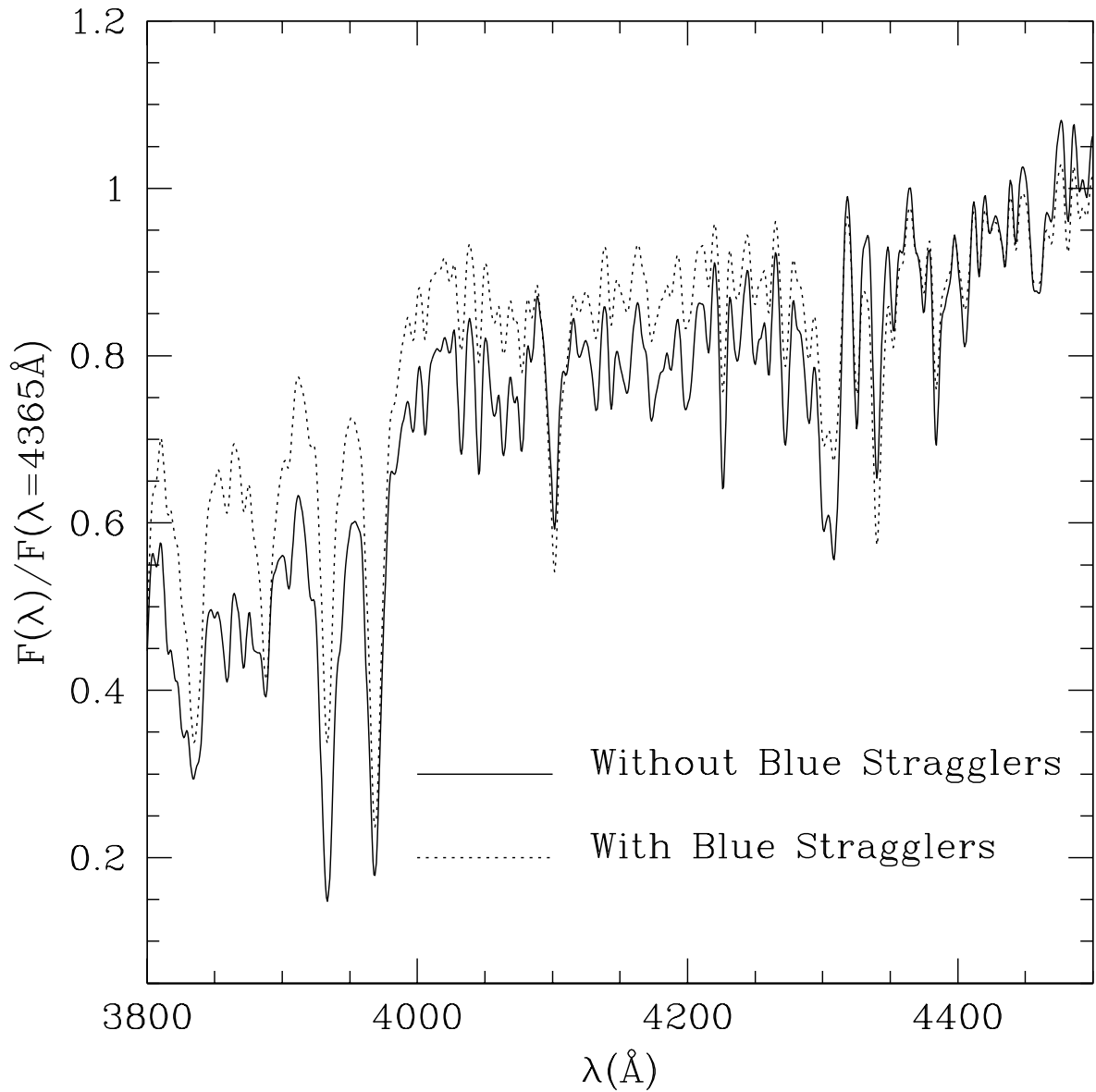}
\caption{Integrated spectrum of M67, resulting from the computations
described in Section \ref{comput}. The spectrum represented by the
solid line was obtained
without considering the contribution due to the blue stragglers. The
spectrum obtained when the latter are included is shown as the dotted
line. The contribution due to the blue stragglers is very large. This is a
robust result, since the number of blue stragglers relative to main sequence
stars is solidly constrained by photometry and membership studies.
} 
\label{specbs}
\end{figure}

The resulting integrated spectrum of M67 is displayed in Figure
\ref{specbs}, where we compare the spectra obtained with and without the
contribution by blue stragglers (BS). As can be seen in this figure,
the contribution to the integrated light due to BS is very large:
when BS are included in the computation, the resulting spectrum is
considerably bluer and has stronger Balmer lines and weaker metal
lines. This result agrees with the finding by Deng et al. (1999),
who modelled the cluster's integrated light using intermediate band
photometry and synthetic stellar spectra from the Kurucz spectral library.
Landsman et al. (1998) assessed the relative contribution of BS stars to
the integrated light of M67 in the ultraviolet and found that BS dominate
the light of the cluster at $\sim 1500 {\rm\AA}$.  It is important to
stress that this result does not depend on any assumption as to the
number ratio of BS to main sequence stars, as the latter number is
anchored on photometry and membership studies. Recall that our recipe
to weigh BS in number is based on the photometry by Fan et al., which
is complete down to many magnitudes below the turnoff.

The BS frequency in M67 appears to be unusually high when compared with
other clusters of similar age and richness (e.g., Ahumada \& Lapasset 1995,
see also a discussion in Landsman et al. 1998).  Hurley et al. (2001),
using binary population synthesis and N-body simulations, propose that
the high BS fraction in M67 is due to a combination of effects related
to binary evolution, stellar encounters in the cluster high density
environment, and evaporation of low-mass stars.

Irrespective of the physical origin of the BS phenomenon, their mere
existence is a potential source of confusion for those attempting to
use Balmer lines to date stellar populations, as is clearly illustrated
by the strength of Balmer lines when BS are included in the integrated
spectrum (Figure \ref{specbs}). In fact, we show in Section \ref{compsp}
that the age one would infer from the integrated spectrum of M67 when
blue stragglers are included is substantially lower ($\sim$ 1.2 Gyr).
However, the case of M67 seems to be exceedingly unusual. As noted by
Landsman et al. (1998), the central stellar density in the cluster is
higher than that found in the centers of most nearby galaxies.

There is another reason to believe that such a high BS contribution does
not seem to be a common phenomenon in the Universe, and in particular
in early-type galaxies, which are the ultimate targets of our modelling
efforts. When blue stragglers are included, the integrated spectrum of
M67 has strong enough Balmer lines that it would be classified as a {\it
k+a/a+k} spectrum according to the definition of several different groups
(Fisher et al. 1998, Balogh et al. 1999, Tran et al. 2003, Poggianti et
al. 2003). Therefore, it is reasonable to assume that such a high fraction
of blue stragglers as found in M67, if common to the centers of early-type
galaxies, would generate a high fraction of {\it k+a/a+k} galaxies in the
Universe. However, recent estimates from the Sloan Digital Sky Survey rate
the fraction of {\it k+a/a+k} galaxies in the local Universe at slightly
lower than 0.1\% (Goto et al. 2003). We therefore conclude that a
blue-straggler population as seen in M67 is, at most, extremely rare in
field galaxies, and henceforth concentrate on comparing our models with
the integrated spectrum of M67 without the contribution of blue
stragglers. This is the topic of the next section.

The spectra shown in Figure \ref{specbs} are helpful templates for the
calibration of stellar population models, in an age/metallicity regime
where these models still remain largely untested. Therefore, we make
them available electronically upon request to the authors.

\section{Comparing Stellar Population Models with the Data for M67}

\subsection{The Turn-off Age of M67} \label{cmdage}

In this section we estimate the age of M67 from the comparison of its
color-magnitude diagram with theoretical isochrones. This is a very
important step, as in the next section we will require our SP computations
to be consistent with the ages inferred from the CMD, using the {\it same}
set of isochrones in both procedures.

The iron abundance of M67 is about or slightly below solar (Brown
1987, Hobbs \& Thorburn 1991, Friel \& Boesgaard 1992, Tautvaisiene et
al. 2000, Shetrone \& Sandquist 2000), and the abundance ratio between
$\alpha$-elements and iron also seems to be nearly solar (Friel \&
Boesgaard 1992, Shetrone \& Sandquist 2000, Tautvaisiene et al. 2000).
Therefore we adopt the isochrones by Girardi et al. (2000) for solar
metallicity.  Girardi et al.'s isochrones are computed for the solar
mixture of heavy elements, and we will henceforth refer to them as the
Padova isochrones.

\begin{figure}
\plotone{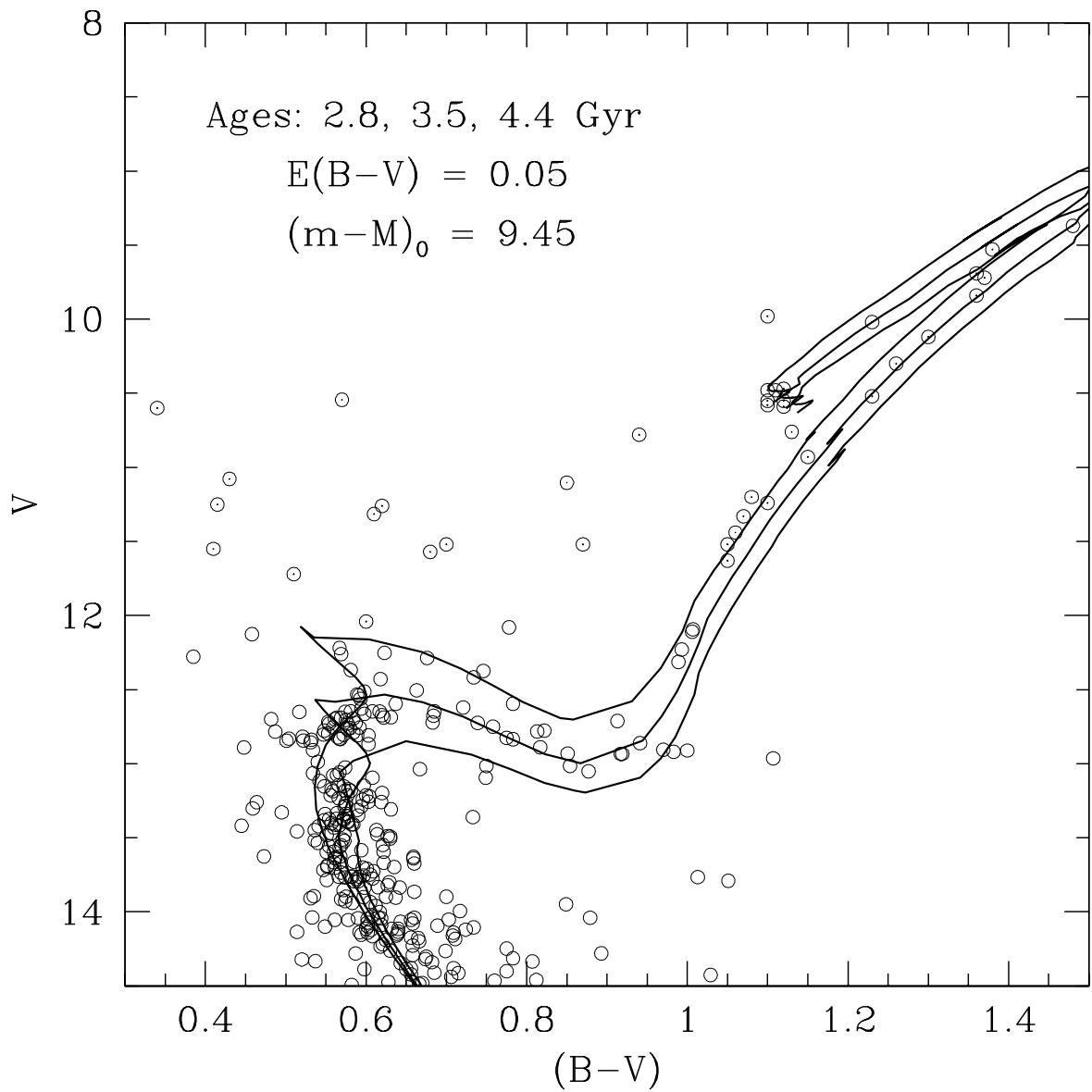}
\caption{Comparison between the color-magnitude diagram of M67 (data
from Montgomery et al. 1993) and the isochrones from Girardi et al.
(2000). The isochrones were transformed to the observational plane using
the recipe described by Schiavon et al. (2002a) and the reddening and distance
modulus indicated in the top left. From left to right, the ages of the
theoretical isochrones are 2.8, 3.5 and 4.4 Gyr.  The age that best
fits the cluster data is 3.5 Gyr.
} 
\label{cmd}
\end{figure}

In Figure \ref{cmd} we display the color-magnitude data for M67 from
Montgomery et al. (1993), overplotted on the Padova isochrones for solar
metallicity and ages of 2.8, 3.5 and 4.4 Gyr. The isochrones were
transformed to the observational plane adopting the color vs. $T_{eff}$
calibration described in Schiavon et al. (2002a) and the bolometric
correction vs $T_{eff}$ calibration from Alonso et al. (1995,1999).

The reddening determinations for M67 range from $E(B-V)=0.03$ (Fan
et al.  1996, Nissen, Twarog \& Crawford 1987) to 0.05 (Montgomery
et al. 1993, Taylor 1980). We adopted a reddening of $E(B-V) = 0.05$,
within the range of previous determinations. The distance modulus adopted
was $(m-M)_0$=9.45 (Chaboyer et al. 1999). The age error due to the
uncertainty in the reddening towards the cluster is very small, mostly
because at ages around 3-4 Gyr the color and luminosity of the turnoff
are very age-sensitive -- as opposed to what happens for older ages
(see discussion in Schiavon et al. 2002b). However, the uncertainties
on the distance modulus and reddening do have some impact on the final
integrated spectrum, as discussed in Section \ref{comput}.

Deciding the exact position of the cluster's turnoff in Figure \ref{cmd}
is not obvious. There is a sparse group of stars between $ V \sim 12.5$
and $\sim 12.0$ at $(B-V) \sim 0.6$ that seems to be well fitted by the
``hook'' above the turnoff of the 2.8 Gyr isochrone. If these stars are
not considered to be blue stragglers, then the best-fitting cluster age
would be $\sim$ 2.8 Gyr. Conversely, if the ``hook'' is considered to be
the clump of stars at $V \sim 12.7$ and $(B-V) \sim 0.6$, the 3.5 Gyr
isochrone would be a better fit. For consistency, we adopt the older
age as the best fit, because the stars brighter than $V \sim 12.5$
were considered to be blue stragglers when we computed the integrated
spectrum of the cluster (see Figure \ref{targets}). Therefore, for
our present purposes, we adopt an age of 3.5 Gyr for M67. We note that
Chaboyer, Green \& Liebert (1999), analysing the same photometric data,
adopted a similar criterion to define the cluster turnoff, and find the
same age for the cluster, even though they adopted a different set of
isochrones and a different distance modulus.  Most importantly, our age
and distance modulus estimates are within the range of previous results
in the literature: age = 3.5 -- 5 Gyr, $(m-M)_0$ = 9.35 -- 9.60 mag (see
the review by Chaboyer et al. 1999 for details). 

The assumption of convective core overshooting can also potentially
affect age determinations in the intermediate age regime ($\simless$
4 Gyr), where turn-off stars have masses $m \simgreat 1.5 M_\odot$
(e.g. Maeder \& Meynet 1991). The isochrones by Girardi et al. (2000)
include mild overshooting for stars with $m > 1.5 M_\odot$ and a gradual
decrease of the overshoot efficiency with decreasing mass down to $m =
1 M_\odot$, where overshooting is turned off. The best-fitting isochrone
in Figure \ref{cmd} has a turn-off mass of $\sim 1.4 M_\odot$, which is
within the regime of low overshooting efficiency in the Girardi et al.
isochrones. Therefore, the age derived here should not be significantly
dependent on the model prescription for overshooting.

Finally we would like to stress that, for the purposes of this paper, it
is not crucial to determine the absolute age of M67. Much more important
for us is to focus on reaching consistency between spectroscopic and
CMD-based ages. This is the topic of the next section.

\subsection{Comparison with Stellar Population Models} \label{compsp}

\subsubsection{The Models} \label{themod}

Before discussing the estimation of the age and metal abundances of
M67 from its integrated spectrum, we recapitulate our modelling
techniques.

The key ingredients of our models are a set of theoretical isochrones from
the literature, a set of calibrations used to transform the latter into
the observational plane, and a set of fitting functions, which are used
to compute absorption line indices as a function of stellar parameters.
The latter are integrated along the isochrones in order to produce
integrated line indices of single stellar populations. The isochrones
adopted here are the ones by Girardi et al. (2000), which were already
presented in Section \ref{cmdage}. The calibrations used to transform
$T_{eff}$ and $M_{bol}$ into colors and magnitudes are a combination of those
presented in Schiavon et al. (2002a) and the empirical calibrations by
Alonso et al. (1995,1999).

The fitting functions are the major new ingredient of our models.
They are based on a redefinition of the Lick/IDS system, rooted in
the Jones (1999) spectral library (a brief description can be found
in Jones \& Worthey 1995), converted to the Lick/IDS resolution,
and supplemented by further data from the original Lick/IDS library
(Worthey et al. 1994). Most importantly, the fitting functions are
based on a new set of homogeneous stellar parameters for the stars
from the Jones spectral library. A detailed description of our stellar
parameter determinations can be found in Schiavon et al. (2002a).  A full
description of the models, including a thorough comparison of our new
fitting functions with those from the literature, will be presented
elsewhere (Schiavon 2003, in preparation).

\subsubsection{The Spectroscopic Age of M67 According to the
SP Models} \label{agesect}

The absorption line indices measured in the integrated spectrum of M67
are listed in Table \ref{tbl-ind} and compared with the predictions of
single stellar population models in Figures \ref {ind1} and \ref{ind2}.
The 1-$\sigma$ errorbars listed in Table \ref{tbl-ind} and adopted in all
the figures involving integrated indices were estimated from the {\it
r.m.s}. scatter between our measurements of Lick indices in Lick/IDS
standard stars and the standard values (see Figure \ref{calib}).
Assuming equal contribution to the {\it r.m.s.} scatter from our
measurements and the standard values, we divide the {\it r.m.s}. by
$\sqrt{2}$ to obtain the error estimates.  These errors are certainly
larger than expected from Poisson statistics and wavelength calibration
errors, and so they are dominated by systematic effects of tying into the
Lick/IDS system with spectra taken from different sources. Since the M32
and M67 composite spectra have very high S/N ratio, comparable to those
of the standard stars, we assume the same overall errors for M32 and M67.

We choose to compare the models with the observations using four age
indicators ($H\delta_F$, G4300, $H\gamma_F$ and $H\beta$) and five
metal abundance indicators (CN1, CN2, Fe4383, $<Fe>$ and Mg $b$). Index
definitions can be found in Worthey et al. (1994) and Worthey \& Ottaviani
(1997). In the figures, dotted lines connect models with the same age,
and solid lines connect models with the same metallicity. The ages
plotted are, from top to bottom, 1.2, 2.0, 2.5, 3.5, 7.9 and 14.1 Gyr,
and the metallicities are, from right to left, [Fe/H] = +0.2, 0.0, --0.4,
--0.7 and --1.3 (the latter two are absent in some plots, for clarity).
The lines connecting models with [Fe/H]=0 and with $t = 3.5$ Gyr are
thicker. Indices measured in the spectra of M67 with and without blue
stragglers are represented by the open square and circle, respectively.

\begin{deluxetable}{cccccccccccccc}
\rotate
\tablecaption{Absorption Line Indices Measured \label{tbl-ind}}
\tablewidth{0pt}
\tablehead{
\colhead {}& $H\delta_F$ &  $H\delta_A$ &  CN1   &    CN2    &  G4300 & 
           $H\gamma_F$ &  $H\gamma_A$ & Fe4383 &  $H\beta$ &  $Mg2$ & 
             Mgb     &    Fe5270    &  Fe5335 } 
\startdata
M67 (no BS) & 0.92 & -0.65 & 0.001 & 0.030 & 4.8 & -0.33 &  -4.0  & 
3.9 & 2.37 & 0.18 & 2.9 & 2.53 & 2.20 \\
M67 (with BS) & 3.05 & 3.31 & -0.078 & -0.045 & 2.6 & 2.16 &  0.7  & 
2.4 & 3.46 & 0.14 & 2.3 & 2.06 & 1.81 \\
M32 & 0.75 & -1.12 & 0.015 & 0.046 & 4.9 & -0.56  & -4.5 &
4.6  & 2.25 & 0.21 & 3.0 & 2.95 & 2.60 \\
Error & 0.09 & 0.14 & 0.005 & 0.008 & 0.1 & 0.05 & 0.2 & 0.2 & 0.09 &
0.01 & 0.1 & 0.06 & 0.08 \\
\enddata
\end{deluxetable}

The most striking feature in these figures is that all Balmer lines
indicate the same age for M67, within less than 1 Gyr. Moreover, the
spectroscopic age according to the Balmer lines is $\sim$ 3.5 Gyr,
in agreement with that inferred in Section \ref{cmdage},
from the comparison of cluster data with theoretical isochrones in
the color-magnitude diagram. We highlight the fact that the {\it
same} isochrones were employed in deriving the spectroscopic and CMD
ages. Even though such consistency should in principle be expected, we
note that previous studies found huge discrepancies between CMD-based
and spectroscopic ages for a well-known Galactic globular cluster,
47 Tuc. The turnoff age of this cluster is 11 $\pm$ 1 Gyr when the
isochrones of Salaris and collaborators (see description in Vazdekis
et al. 2001) are employed.  However, Gibson et al. (1999) found a
spectroscopic age in excess of 20 Gyr for the cluster.  Later on,
Vazdekis et al. (2001) showed that better agreement was reached when
effects like $\alpha$-enhancement and diffusion of heavy elements are
taken into account. However, a mismatch of 4 Gyr was still lingering.
Schiavon et al. (2002b) finally showed that most of the discrepancy was
due to a mismatch between the theoretical luminosity function of the
brighter giant stars and the observations, in the sense that theory
predicted too few bright giants. Therefore, considering the previous
history, it is quite reassuring that our single stellar population models
provide results at this level of consistency, in a regime where SP models
have never been tested before in such a level of detail.

The remaining age indicator, the G band (G4300) presents a trend as a
function of age which is opposite to that of Balmer lines: it gets weaker
for younger ages. This is due to the fact that it is much stronger in
giants than in turnoff stars. As a result, the G band gets stronger in
the integrated spectra of older stellar populations because they are
characterized by a higher relative contribution of giant stars to the
integrated light. Being mostly due to the CH molecule, the G band is
not expected to be a clean age indicator, due to its dependence on the
abundance of carbon. In fact, we obtain an older age for M67 from the EW
of the G band ($\sim$ 6 Gyr).  Tripicco \& Bell (1995) showed that the G
band is mostly sensitive to the abundance of carbon.  According to their
calculations, the mismatch between the model for [Fe/H]=0.0 and age =
3.5 Gyr is such that it could be explained by M67 having a higher than
solar [C/Fe]. However, abundance studies of M67 turnoff stars point
at a different direction: they have nearly solar [C/Fe], and may even
perhaps be a little carbon-poor (see discussion in Section \ref{metsect}).

Therefore, it is quite puzzling that in the M67 integrated spectrum the G
band is stronger than in the model for 3.5 Gyr and solar metallicity. We
tried removing stars very deviant from the average relation between the
EW of the G band and color, but that did not remove the discrepancy.
We showed in Section \ref{conver} that the calibration of our G band
measurements into the Lick/IDS system was somewhat uncertain. We thus
believe that calibration uncertainties related to the conversion into
the Lick/IDS equivalent width system are responsible for this mismatch.

In Section \ref{contm}, we corrected our model predictions for the effect
of AGB stars, by adding an extra 0.1 dex in the luminosity function of
the upper giant branch.  In their analysis of the integrated spectrum
of 47 Tuc, Schiavon et al. (2002b) showed that an error of $\pm$ 0.1
dex in the luminosity function of giants brighter than the horizontal
branch translates into an error of $\pm$ 1 Gyr in the spectroscopic age
inferred from Balmer lines. The latter result is valid for an old stellar
population ($\sim$ 11 Gyr) with [Fe/H]=--0.7. For a younger and more
metal-rich stellar population, we expect the effect to be less important,
given that the total contribution by cool giants to the optical integrated
light of metal-rich SPs with intermediate ages is lower. In fact, we note
that the integrated spectrum of M67 is little affected by uncertaintities
in the LF of the upper giant branch: the change in the M67 indices in
Figures \ref{ind1} and \ref{ind2} that would be generated by a 0.1 dex
variation in the LF above the horizontal branch is in all cases smaller
than the symbol size.

In summary, we achieve excellent consistency for M67 between the
spectroscopic ages inferred from all Balmer lines studied.  We attribute
the mismatch with the spectroscopic age inferred from the G band to
uncertainties in the calibration of the latter.

\subsubsection{The Metal Abundances of M67 According to the SP Models}
\label{metsect}

\begin{figure}
\plotone{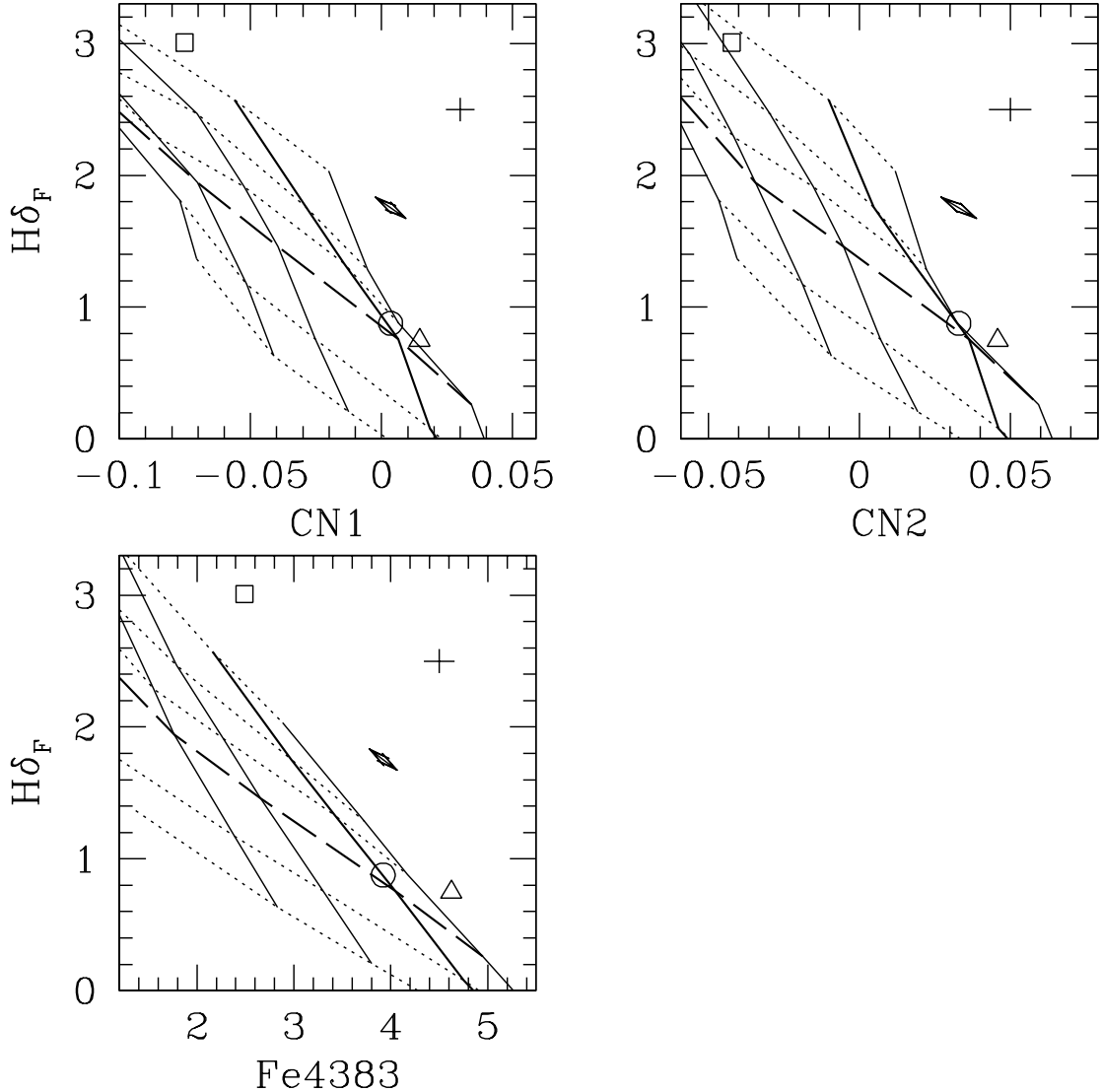}
\caption{Comparison of model predictions absorption line indices for
single stellar populations with those measured in the spectra of M32
(open triangle), M67 without blue stragglers (open circle) and M67
including blue stragglers (open square) Dotted lines connect model
predictions for the same age. From top to bottom: 1.2 2.0, 2.5, 3.5,
7.9, and 14.1 Gyr. Solid lines connect model predictions for the same
[Fe/H]. From right to left: +0.2, 0.0, --0.4, --0.7, and --1.3 (in some
c, the line for [Fe/H]=--1.3 falls outside the plotting area). The lines
for [Fe/H]=0.0 and age=3.5 Gyr are thicker, for clarity. The arrows
indicate by how much the integrated indices of M67 vary for an error
of $\pm$ 0.015 in $E(B-V)$. A higher reddening makes Balmer lines look
stronger and metal lines weaker.  The spectroscopic age of M67, when
blue stragglers are not included in the integrates spectrum, is about 3.5
Gyr in all diagrams, which is in agreement with the value inferred from
the CMD. The data on M32 indicate that it has super-solar iron abundance
and has a nearly solar abundance of carbon and/or nitrogen, relative 
to iron.
}
\label{ind1}
\end{figure}

\begin{figure}
\plotone{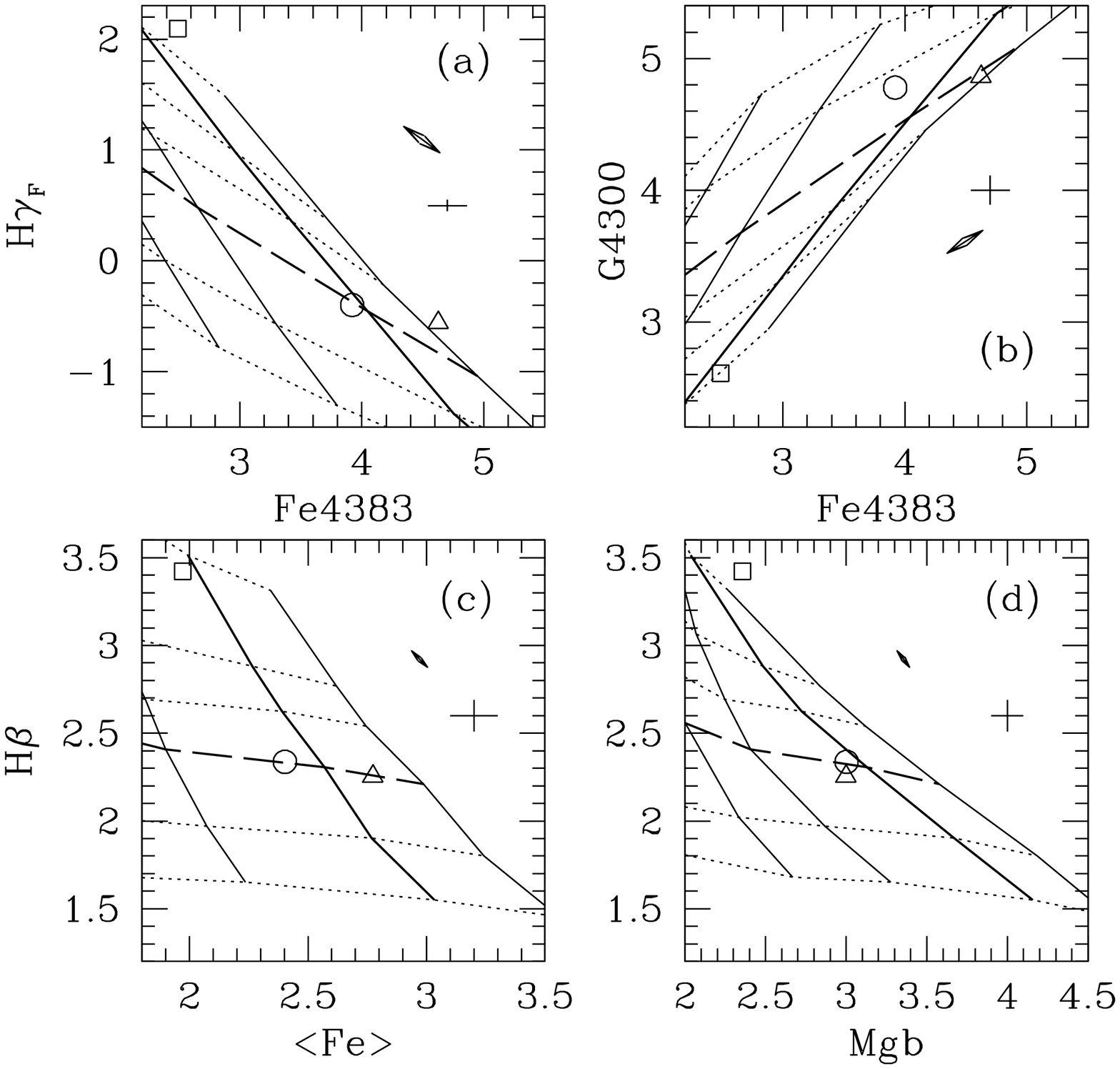}
\caption{Same as Figure \ref{ind1}. Symbols are the same as for Figure
\ref{ind1}. The most important conclusions to be
drawn from this figure and Figure \ref{ind1} are: 1) All Balmer lines
indicate the same age for M67, in agreement with the CMD-based age;
2) The Fe, CN and Mg indices indicate solar abundances within $\pm$
0.1 dex, in agreement with spectroscopic studies of stellar members.
3) When blue stragglers are included in the integrated spectrum of
M67, the spectroscopic age drops to $\sim$ 1.2 Gyr, or even younger,
depending on the line index adopted.  For the G band the dependence on
age is inverted, so that age decreases from top (14.1 Gyr) to bottom
(2.0 Gyr). Ages based on the G band, are too old (a factor of two in the
case of M67). For M32, bluer Balmer lines tend to indicate younger ages,
and bluer Fe lines indicate higher metallicities. The data on M32 indicate
that it has super-solar iron abundance, and is underabundant in magnesium,
relative to iron. See discussion in the text.
} 
\label{ind2}
\end{figure}

Figures \ref{ind1} and \ref{ind2} also allow checking the consistency
of SP model predictions with the known elemental abundances of M67.
This is a unique opportunity, because M67 stars have been the subject of
many abundance analyses by different groups, so that it can be fairly
said that the abundance pattern of the cluster is well known. The
importance of enforcing consistency between SP models and the results
of classical stellar abundance analyses lies in the fact that the latter
provide the only physically meaningful measurement of the abundances of
heavy elements in stars. We emphasize that such a tight consistency is
seldom attempted in the literature. A similar approach was followed in
our study of the mildly metal-rich Galactic globular cluster, 47 Tuc
(Schiavon et al.  2002a,b), and now we provide, for the first time,
a check of the consistency of SP models in the regime of intermediate
age and solar metallicity.  For this purpose, we below discuss separately
a few relevant elements, together with a brief summary of the previous
results from detailed abundance analysis of M67 stars.

\subsubsubsection{Iron}

Classical abundance analysis of M67 members, based on high-dispersion
spectroscopy, was pioneered by Cohen (1980), who found $<$[Fe/H]$>$ =
--0.39, from a study of four cluster giants.  This result was revised by
subsequent studies which found a higher iron abundance, near the solar
value. Foy \& Proust (1981), analysing two giant stars, found $<$[Fe/H]$>$
--0.1 $\pm$ 0.1. Garcia Lopez, Rebolo \& Beckman (1988), who studied seven
turnoff and one giant star found $<$[Fe/H]$>$ = +0.04 $\pm$ 0.04.  Friel \&
Boesgaard (1992) analysed three cluster dwarfs and found $<$[Fe/H]$>$ =
+0.02 $\pm$ 0.1. Most recently, Tautvaisiene et al. (2000) found $<$[Fe/H]$>$
= --0.03 $\pm$ 0.15, in an analysis of nine giant stars, and Shetrone and
Sandquist (2000) determined $<$[Fe/H]$>$ = --0.05 $\pm$ 0.05 for four turnoff
stars. In view of these previous studies, we assume that M67 has a solar
iron abundance ($<$[Fe/H]$>$ = 0 $\pm$ 0.1).

The line indices in Figures \ref{ind1} and \ref{ind2} that are most
sensitive to the iron abundance are $<Fe>$ and Fe4383 Figures \ref{ind1}
and \ref{ind2} indicate that both indices are consistent with $<$[Fe/H]$>$
= 0 to within less than 0.1 dex.

\subsubsubsection{Carbon and Nitrogen}

The abundances of these elements are very important because they
affect significantly the indices blueward of 4500 ${\rm\AA}$, due to
the presence of thousands of CN and CH lines (for a discussion, see
Schiavon et al. 2002a). The abundance of carbon also has an important
effect on Mg $b$, due to contamination by $C_2$ lines (Tripicco \& Bell
1995). Brown (1987), analysing stars at different evolutionary stages,
found that the C/N abundance ratio in M67 stars undergoes a sharp
drop between $M_V \sim 3.5$ and $\sim 2.8$, as a result of the first
dredge-up episode. Subsequent analyses are consistent with this result.
Regarding carbon, analyses of dwarf stars by Friel \& Boesgaard (1992)
and Shetrone \& Sandquist found $<$[C/Fe]$>$ = --0.1 $\pm$ 0.2 and $0.0 \pm
0.1$, respectively, while Tautvaisiene et al. (2000), analysing giant
stars, found $<$[C/Fe]$>$ = --0.2. Nitrogen abundances are more uncertain,
and also there are fewer determinations in the literature. Tautvaisiene
et al.  found that M67 giants have $<$[N/Fe]$>$ = +0.2, whereas Shetrone
\& Sandquist, analysing turnoff stars, found $<$[N/Fe]$>$ = +0.1 $\pm$ 0.2.
In summary, [C/N] is $\sim$ 0.25 dex lower in M67 giants than in dwarfs.
In particular, we emphasize that the abundance ratios in M67 giants
are nonsolar, whereas M67 dwarfs are consistent with a solar
carbon abundance, while perhaps being slighty enhanced in nitrogen.

In Figure \ref{ind1}, it can be seen that our model prediction for 3.5 Gyr
and [Fe/H] = 0 is in excelent agreement with the measurements of the CN
indices in the integrated spectrum of M67, even though our models have,
nominally, solar abundance ratios. This can be understood in view of the
results of Gratton et al. (2000) and Carretta, Gratton \& Sneden (2000),
who showed that in field stars the first dredge-up operates a change in
the surface composition of carbon and nitrogen which is very similar to
what is seen in M67 and other clusters.  Therefore, in the field stars
that serve as the basis of our models, the run of C,N abundances as a
function of evolutionary stage is arguably similar to that of M67 stars,
so that the effects due to stellar evolution in M67 stars are accounted
for in the models by construction (this has first been pointed out by
Gorgas et al. 1993). Therefore, no correction due to nonsolar abundance
ratios needs to be applied to our model predictions and M67 stars can be
considered to constitute a single stellar population with solar [C,N/Fe].

\subsubsubsection{Magnesium}

The index in Figure \ref{ind2} which is most sensitive to the abundance
of magnesium is Mg $b$, which is the most popular estimator of the
metallicities of stellar populations of galaxies, and therefore it is
very important that our predictions be consistent with the abundance
analyses of M67 stars for magnesium. There is some disagreement in the
literature as to the magnesium abundance of M67.  Foy \& Proust found
$<$[Mg/Fe]$>$ = 0, Tautvaisiene et al. (2000) found $<$[Mg/Fe]$>$ =
+0.1, and Shetrone \& Sandquist found $<$[Mg/Fe]$>$ = --0.1. In Figure
\ref{ind2}, it can be seen that our model prediction for Mg $b$ is
somewhere between [Mg/Fe] = --0.1 and 0.0, and so it is consistent with
abundance analysis work. However, we stress that more work is needed in
order to ascertain the magnesium abundance of M67.

\bigskip
\bigskip
\bigskip

In summary, we conclude that our model predictions are consistent to
within $\pm$ 0.1 dex with the known metal abundances of M67.  This is
reassuring, because it implies that our models predict the correct
values of abundance-sensitive indices which are among the most widely
used in stellar population work. We also emphasize that the predictions
are consistent for indices which are sensitive to the abundances of a
number of important chemical species, like iron and the light elements
carbon and nitrogen, and magnesium. This also means that the models can
be used to estimate luminosity-weighted abundance ratios of galaxies in
a way that is consistent with the state of the art of our knowledge of
stellar abundances in the Galaxy.

\section{The Spectroscopic Age and Metal Abundances of M32}
\label{m32}

We would like to start our discussion of the SP properties of M32 by
reminding the reader that our spectrum is integrated within a square
3'' on a side positioned on the center of the galaxy. This is important
in view of the fact that previous studies have indicated the presence
of a radial gradient in the spectroscopic properties of M32 (Davidge
1991, Gonz\'alez 1993, Hardy et al. 1994, Worthey 2003).

\begin{figure}
\plotone{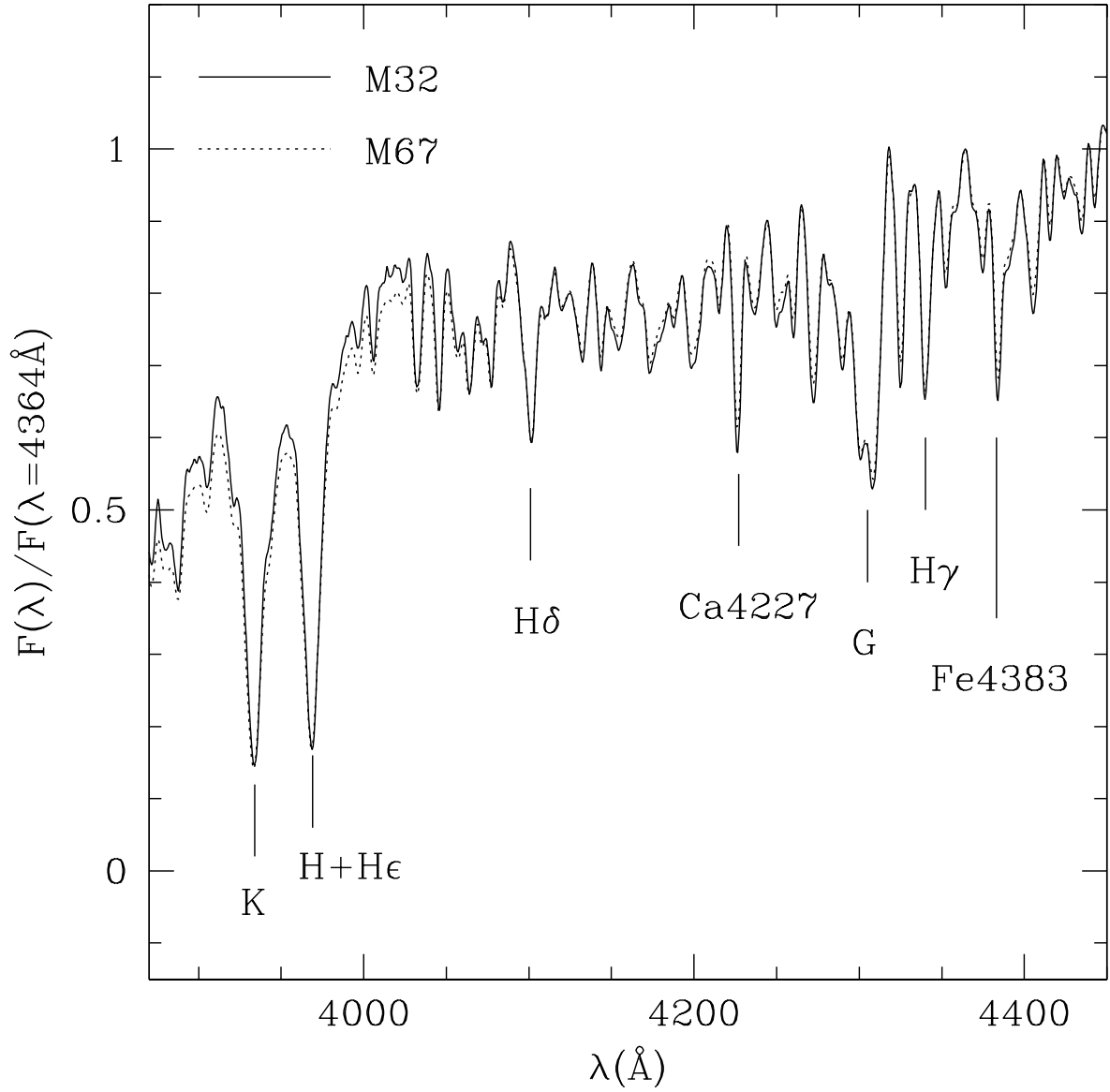}
\caption{Comparison between the integrated spectra of M67 and M32. The
spectra are very similar, although M32 looks to be slightly more
strong-lined than M67, which suggests that M32 is on average slightly more
metal-rich and younger than M67. See discussion in Section
\ref{m32}.
} 
\label{specms}
\end{figure}

Because of its morphology, proximity, and very high central surface
brightness, M32 has become a benchmark for the study of stellar
populations, and galaxy evolution in general. Spinrad \& Taylor
(1971), and Faber (1973) pioneered the efforts to determine the
luminosity-weighted mean metallicity of the central parts of M32,
from low-resolution spectrophotometric observations, but O'Connell
(1980) was the first to find that such data for the center of M32 could
only be matched by requiring an intermediate age population ($\sim$
5 Gyr) and near-solar metallicity. More recent attempts, based on more
sophisticated methods and/or modelling, obtained similar results, with
age ranging between 3 and 5 Gyr and [Fe/H] within 0.1 dex from the solar
value (e.g. Rose 1994, Jones \& Worthey 1995, Vazdekis \& Arimoto 1999,
Trager et al. 2000, Worthey 2003, Caldwell et al. 2003). These results are
confirmed by Figure \ref{specms}, where we compare the bluer part of the
integrated spectrum of M67 that was constructed in Section \ref {comput}
with the observed integrated spectrum of M32.  Some of the most important
spectral indices employed in our analysis are indicated in the Figure. As
can be seen, the two spectra are very similar. Actually, the integrated
spectrum of M32 is in general slightly more strong-lined than that of M67,
which suggests a slightly younger age and a slightly higher metallicity.

The latter result can be inspected more quantitatively in Figures
\ref{ind1} and \ref{ind2}, where Lick indices are compared with
predictions for single SPs. In order to minimize abundance ratio effects,
we first focus on the plots of Balmer lines against Fe indices. According
to these plots, the spectroscopic age of M32 is somewhere between 2.0 and
3.5 Gyr, whereas its luminosity-weighted iron abundance is super-solar,
with +0.1 $\simless$ [Fe/H] $\simless$ +0.3. These numbers confirm our
expectations based on inspection of the spectra of Figure \ref{specms}.

It is important to bear in mind that the strength of Balmer lines is
not solely governed by the temperature and brightness of turnoff stars:
other families of hot stars, such as blue stragglers and hot horizontal
branch (HB) stars can affect Balmer line strengths, thus hindering their
use as age indicators (e.g. Freitas Pacheco \& Barbuy 1995, Lee, Yoon \&
Lee 2000, Maraston et al. 2003). In principle, a small population of hot
stars could play a major role in strengthening the Balmer lines in M32,
thereby reducing, or eliminating, the necessity for the $\sim$2-3 Gyr
old mean population implied by the Balmer line strengths.  However,
Caldwell et al. (2003) have demonstrated that to explain the enhanced
Balmer lines in M32 requires a $\sim$30\% contribution from a hot-star
component, whether from a blue HB or blue stragglers.  On the other hand,
they also show that the Ca II index, which levers the strength of Ca II K
versus H$\epsilon$ + Ca II H, restricts the contribution from hot stars
to a level of $\sim$7\%.  In addition, M32 has the weakest UV upturn in
the sample of Burstein et al. (1988), so that its relative number of
hot HB stars must be low.  In principle, cooler BS stars, that do not
make such a strong impact on the above-mentioned Ca II index, could be
present in M32, but then they would have to exist in exceedingly large
numbers, in fact rivalling the numbers of true MSTO stars.  In short, we
have reasons to be confident that our Balmer line measurements primarily
provide an estimate of the spectroscopic age of M32.

With the above caveats in mind, we now focus on the plots of Balmer lines
against CN and Mg indices, in order to assess the abundance ratios of M32.
The Mg $b$ index is mostly sensitive to magnesium. In panels (c)
and (d) of Figure \ref{ind2} it is seen that M32 has [Fe/H] $\sim$
+0.1 according to $<Fe>$, while, according to Mg $b$ it is about 0.1
dex below the solar metallicity model, which is strongly suggestive
of [Mg/Fe] $<$ 0.  In the case of the CN indices (panels a and b of
Figure \ref{ind1}), in both cases the M32 data fall nearly on top of the 
[Fe/H]=+0.2 model. Since both CN1 and CN2 are mostly sensitive to the
combined effect of nitrogen and carbon abundances, and given the above
discussion on the iron abundance of M32, it is likely that M32 has a
nearly solar abundance of carbon and/or nitrogen relative to iron. 

\begin{figure}
\plotone{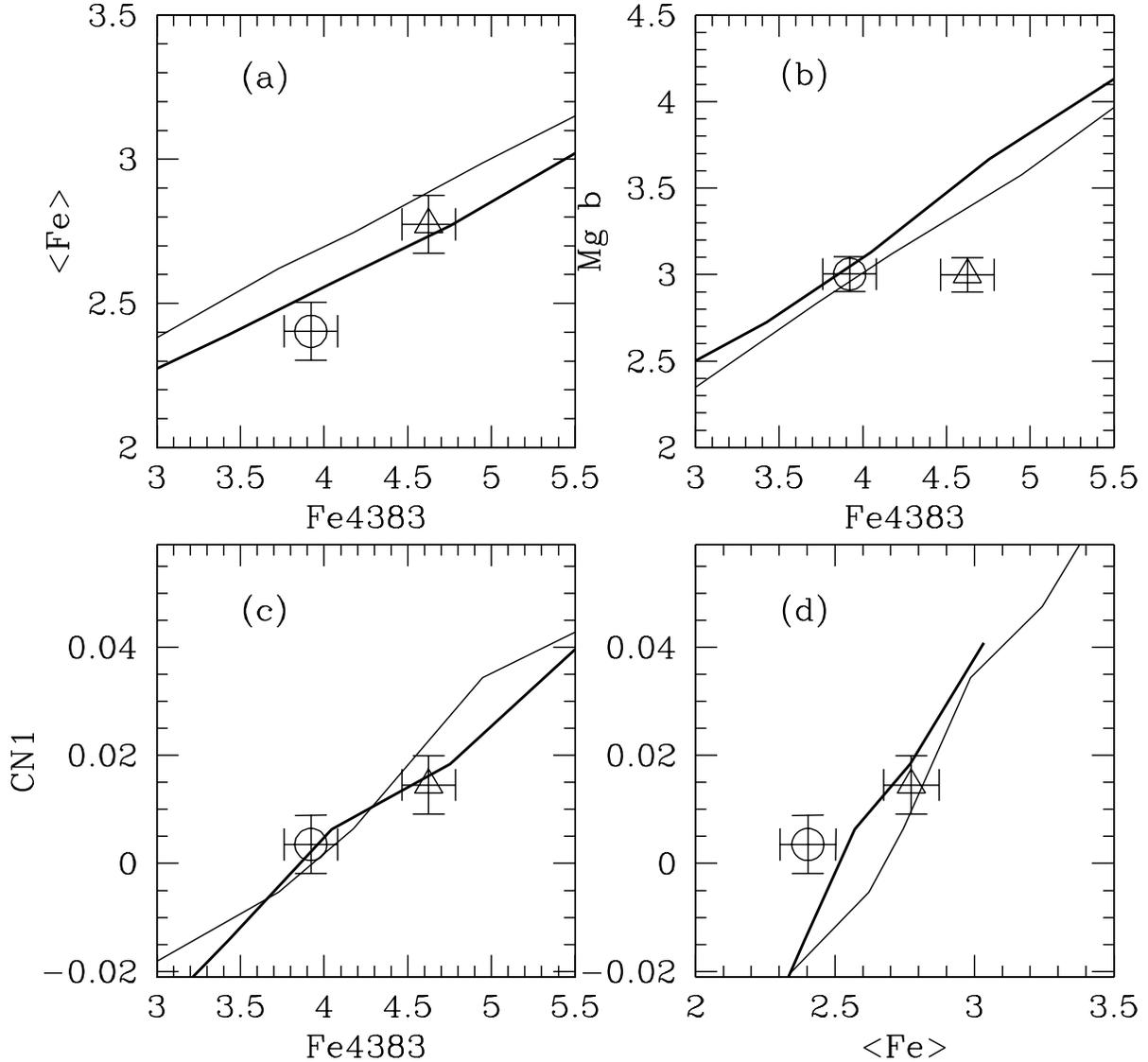}
\caption{The data on M67 and M32 are compared to single SP models on
plots involving only abundance-sensitive indices. The symbols adopted
are the same as in Figure \ref{ind1}. Such plots are useful to unveil
nonsolar abundance ratios. Again, the open triangle and circle represent
M67 and M32, respectively. The model lines are for [Fe/H] = 0.0 and +0.2.
The former is represented by a thick line. The lines connecting same-age
models are omitted for clarity. While M67 is basically consistent with
solar in all abundance ratios, M32 appears to have [Mg/Fe] substantially
below solar.  The combined abundances of C and N appear to be nearly
solar in the center of M32.  }
\label{ind3}
\end{figure}

Abundance ratios can be better assessed in plots of metal index against
metal index. Such plots are shown in Figure \ref{ind3}, where we compare
the measurements with the model predictions of single stellar populations
for [Fe/H] = 0.0 and +0.2. The lines connecting models with same age
are omitted in these plots for clarity. In panel (b) it can be seen
that while M67 looks like having strictly solar [Mg/Fe], M32 deviates by
roughly 3 $\sigma$ from the model prediction for solar abundance ratios,
which confirms the suggestion above that it has [Mg/Fe] $<$ 0. The case
for nonsolar abundance ratios for carbon and/or nitrogen is perhaps
less convincing, because it depends to some extent on the Fe abundance
indicator adopted. In panel (c), where CN1 is plotted against Fe4383,
M32 is clearly deviating from the solar ratio models, but less so when
CN1 is compared with $<Fe>$ in panel (d) (essentially the same results
are obtained in plots involving CN2). We point out that our results
agree with those of Worthey (2003) as regards [Mg/Fe], but we are in
disagreement when it comes to the abundances of carbon and nitrogen, for
which he found an enhancement relative to iron of $\sim$ 0.2 dex. 

Finally, we would like to point out an apparent systematic trend of the
mean age and metallicity inferred for M32 as a function of wavelength:
the resulting age looks lower, and the metallicity looks higher, as we
go to bluer indices.  According to the $H\beta$ vs. $<Fe>$ plot,
the central parts of M32 have luminosity-weighted mean metallicity and
age given by [Fe/H] $\sim$ +0.1 and $\sim$ 3.5 Gyr. According to the
$H\gamma_F$ vs. Fe4383 plot, [Fe/H] is slightly above +0.2,
and the mean age is $\sim$ 2.5 Gyr, while in the $H\delta_F$ vs.
Fe4383 plot the mean metallicity is even higher and the mean age younger
($\sim$ 2.0 Gyr). Even though the spread in the age determinations is
relatively small (due to the high age sensitivity of Balmer lines in
this age regime), it is significant, given the small errorbars in our
measurements. If we assume that the mean age and [Fe/H] of M32 are
as given by the $H\beta$ vs. $<Fe>$ plot (3.5 Gyr, +0.1 dex),
then the measurements of both $H\gamma_F$ and $H\delta_F$ are off their
predicted positions when plotted against Fe4383 by more than 2 $\sigma$.

This result should be contrasted against the case of M67, for which
we found essentially the same age for all Balmer lines, and iron
abundances according to different Fe indices in agreement to within
0.1 dex. M32 differs from M67 in two basic aspects: 1) It is not a
single stellar population, but rather a mix of stars of different
metallicities (Grillmair et al. 1996) and possibly different ages; 2)
As discussed above, M32 has an abundance pattern that differs from that
of the Sun. 

A simple calculation shows that a very small fraction of a young ($\sim$
1.5 Gyr) stellar population on top of an older one can reproduce data
for Balmer and Fe lines. The younger stellar population is much brighter
in the blue, so that it would affect bluer spectral indices more strongly
with the result that the bluer Balmer lines of such a simple composite
stellar population would indicate younger ages.

On the other hand, it might be also possible to reproduce the data
by applying the method championed by Trager et al. (2000), where the
absorption line indices are corrected for varying abundance ratios,
based on the computations of synthetic stellar spectra by Trippico \&
Bell (1995). The latter does not include calculations for $H\delta_F$
and $H\gamma_F$ though, because these indices were defined later, by
Worthey \& Ottaviani (1997). This is important because we have shown,
in Schiavon et al. (2002a), that both $H\gamma_F$ and $H\delta_F$ are
influenced by variations in the abundances of carbon and nitrogen, due
to the contamination by CN and CH lines. While we cannot correct the
higher order Balmer lines for the effects of unknown abundance ratios,
we recognize that they might be at least partially responsible for the
results of Figures \ref{ind1} and \ref{ind2}.

It is very hard to disentangle the effect of varying abundance ratios
from the potential age mix of stellar populations. As a consequence,
when blue indices are involved, it is difficult to apply the method
devised by Trager et al. (2000) to assess nonsolar abundance ratios.
The reason is that the method essentially seeks the combination of
metal abundances that brings the ages and metallicities into agreement
according to the different line indices, so that it tends to wash out
any effect due to a mix of stellar populations of different ages. In a
future paper, we will address this problem in more detail.

\section{Conclusions} 

We have constructed an integrated spectrum of the Galactic cluster M67,
from spectroscopic observations of bona fide cluster members. The
resulting integrated spectrum was used as a template for stellar
population synthesis, through comparison with model predictions for
single stellar populations. We constrasted the spectroscopic age and
metal abundances estimated from the SP models with the age obtained from
the cluster CMD, and elemental abundances determined in previous works
from high dispersion abundance analyses. It is the first time that such
models are tested to this level of precision in the solar metallicity,
intermediate-age regime. Our SP models were then applied to the study
of the integrated spectrum of M32, yielding reliable determinations of
the galaxy's light-weighted mean age and metal abundances. We summarize
below our main conclusions.

1. Our models for the known age and metal abundances of M67 match
very well its integrated spectrum. Adopting the Girardi et al. (2000)
theoretical isochrones for solar metallicity and solar abundance ratios,
we matched both the color-magnitude data and integrated spectrum of M67,
for an age of 3.5 Gyr and solar metallicity, to within 0.5 Gyr and 0.1
dex respectively. Most importantly, we found very good consistency
for the ages derived from $H\beta$, $H\gamma$ and $H\delta$. This
result demonstrates the remarkable degree of consistency of our models.
We also performed a detailed check to see whether our model predictions
are consistent with the known elemental abundances of M67 stars. We found
that the predictions from SP models for the abundances of iron, carbon,
nitrogen and magnesium agree with the elemental abundance analyses of
individual stars to within 0.1 dex.

2. The integrated spectrum of M32 is very similar to, though slightly more
strong-lined than, that of M67. On the basis of our newly tested models,
we perform an analysis of the integrated spectrum of M32, with the aim
of estimating its light-weighted mean age and metal abundances. We found
that the spectroscopic age of M32 is somewhere between 2 and 3.5 Gyr,
depending on the Balmer line adopted.  For the mean metal abundances,
we found [Fe/H] ranging between +0.1 and +0.4, again depending on the
metallicity indicators adopted. For light elements, we found M32 to be
underabundant in magnesium relative to iron, with [Mg/Fe] $\sim$ --0.2,
whereas the combined abundances of carbon and nitrogen seem to be nearly
solar relative to iron. A more in-depth study of the composite nature
of the stellar populations in M32 is deferred to a forthcoming paper.

3. Contrary to the case of M67, application of our single stellar
population models to the integrated spectrum of the central parts of M32
revealed a systematic effect in the resulting ages and metallicities as a
function of wavelength. Use of red indices results in a mean age $\sim$
1.5 times older and a metallicity $\sim$ 0.3 dex lower than when blue
indices are adopted. This result can be due to either the effects of
nonsolar abundance ratios on blue absorption indices, and/or to a mix
of stellar populations in M32, with different ages.


4. In agreement with previous studies, we found that blue stragglers
contribute with an unusually high fraction of the blue integrated
light of M67. This result is robust, since our integrated spectrum is
constructed on the basis of photometry which is complete down to many
magnitudes below the turnoff. We suggest that this result might be due
to the cluster's past history of severe mass segregation, followed by
evaporation of low-mass stars.

Calibrating models by comparison with integrated spectra of clusters
has become a standard procedure in the field of stellar population
synthesis. The nature of the Galactic globular cluster system, and
the limitations of our knowledge of the detailed abundance pattern
of their constituent stars, has prevented an accurate calibration of
such models with the degree of accuracy that is required for many of
their applications. The latter is particularly true in the regime of
solar metallicity and intermediate ages, which is of great
importance in view of its potential application for the study of
stellar populations of distant systems, for which the lookback times
are more than half the Hubble time. This work is an attempt at
filling this gap. Applications of our models to the study of
galaxies both nearby and at cosmological distances are currently
under way.

\acknowledgments{We would like to thank S. Faber for enlightening
discussions during the course of this investigation. The referee, Guy
Worthey, is thanked for comments and suggestions that greatly improved
this paper. We also thank Hyun-chul Lee for a careful reading of an
earlier version of this manuscript. R.P.S. acknowledges support provided
by the National Science Foundation through grant GF-1002-99 and from the
Association of Universities for Research in Astronomy, Inc., under NSF
cooperative agreement AST 96-13615 (Gemini Fellowship). Support from the
NSF, through grant AST-0071198 to the University of California, Santa Cruz
and from CNPq/Brazil, under grant 200510/99-1 are also deeply thanked.
J.A.R.  acknowledges support from NSF grant AST-9900720 to the University
of North Carolina}

\appendix

\begin{deluxetable}{cccccccccccc}
\rotate
\tablecaption{Lick indices measured in the spectra of 
field and M67 stars
\label{tbl-ews}}
\tablewidth{0pt}
\tablecolumns{4}
\tablehead{
\colhead{ID} & $H\delta_F$ & CN1 & CN2 & G4300 & $H\gamma_F$ & Fe4383 & 
$H\beta$ & Mg $b$ & Mg $2$ & Fe5270 & Fe5335}
\startdata
HD37216   &-0.0927 &-0.0268 &-0.0090 & 5.5800 &-1.7390 & 5.0640 & 1.9310 & 4.0450 & 0.1866 & 2.6840 & 2.3580 \\
HD47914   &-1.3580 & 0.1929 & 0.2470 & 6.0950 &-3.6050 & 8.9290 & 0.6220 & 4.6620 & 0.4279 & 4.3080 & 4.3690 \\
HD49178   & 0.7858 &-0.0070 & 0.0108 & 5.2630 &-0.6413 & 3.7520 & 2.6100 & 3.2390 & 0.1454 & 2.1010 & 1.8940 \\
HD60522   &-1.2280 & 0.1305 & 0.1837 & 5.8620 &-3.2510 & 8.7260 & 0.7104 & 4.7330 & 0.4284 & 4.2310 & 4.3430 \\
HD61606A  &-1.1130 & 0.0038 & 0.0308 & 5.7230 &-3.2210 & 7.5730 & 0.9142 & 5.9370 & 0.3588 & 3.7920 & 3.6200 \\
HD62721   &-0.9038 & 0.1198 & 0.1765 & 6.0490 &-3.0220 & 7.7080 & 0.4400 & 4.8540 & 0.4132 & 3.8200 & 3.7120 \\
HD69582   & 0.4846 &-0.0427 &-0.0294 & 4.9650 &-0.6286 & 4.0670 & 2.4240 & 3.3490 & 0.1475 & 2.4380 & 2.0690 \\
HD69830   &-0.0913 &-0.0208 &-0.0053 & 5.7350 &-1.8680 & 4.9510 & 1.9740 & 4.3250 & 0.1949 & 2.6830 & 2.2900 \\
HD70272   &-1.4280 & 0.1288 & 0.1807 & 5.8610 &-3.0130 & 8.8740 & 0.7205 & 4.2880 & 0.3889 & 4.3060 & 4.3320 \\
HD94705   &-0.6903 &-0.1174 &-0.0824 & 3.5000 &-2.2890 &-0.2740 & 5.0380 &14.0800 & 0.4313 & 3.4070 & 1.2200 \\
HD98991   & 3.3470 &-0.1002 &-0.0778 & 1.5110 & 3.5730 & 0.8280 & 4.3510 & 0.8016 & 0.0480 & 1.1230 & 0.9891 \\
HD106156  &-0.1425 & 0.0264 & 0.0430 & 5.6730 &-1.8050 & 5.4120 & 2.1650 & 4.4290 & 0.2197 & 3.0360 & 2.6080 \\
HD128987  & 0.1842 &-0.0374 &-0.0236 & 4.9530 &-1.0410 & 4.3020 & 2.1810 & 3.7900 & 0.1631 & 2.6530 & 2.2100 \\
HD126511  & 0.0126 & 0.0123 & 0.0278 & 5.4960 &-1.6630 & 4.9920 & 2.3380 & 4.2010 & 0.1986 & 2.5700 & 2.4120 \\
HD136834  &-1.1980 & 0.1275 & 0.1618 & 5.9300 &-3.5640 & 7.9070 & 1.2520 & 6.4510 & 0.4054 & 4.1830 & 3.8950 \\
HD158614  & 0.4254 &-0.0222 &-0.0110 & 5.3790 &-1.1710 & 4.0860 & 2.4000 & 3.5730 & 0.1599 & 2.4140 & 2.1240 \\
HD165341  &-0.4514 & 0.0045 & 0.0196 & 5.6020 &-2.2010 & 5.5210 & 1.7720 & 4.7960 & 0.2407 & 3.2220 & 2.7730 \\
    5041  & 1.8350 &-0.0712 &-0.0505 & 4.1610 & 1.1570 & 2.0860 & 3.1420 & 1.4940 & 0.0735 & 1.7170 & 1.3170 \\
    5059  &-0.6574 & 0.0772 & 0.0985 & 6.4420 &-2.7460 & 5.4150 & 1.6330 & 3.4310 & 0.2000 & 2.9500 & 2.4030 \\
    5118  & 2.5360 &-0.0914 &-0.0688 & 2.9890 & 2.2230 & 1.2170 & 3.5770 & 1.3260 & 0.0758 & 1.3320 & 1.0400 \\
    5169  & 2.1560 &-0.0833 &-0.0597 & 3.6500 & 1.6450 & 1.8200 & 3.4600 & 1.5350 & 0.0795 & 1.5380 & 1.3140 \\
    5191  & 3.0040 &-0.1002 &-0.0786 & 2.2490 & 2.8480 & 1.0120 & 3.8600 & 1.2530 & 0.0623 & 1.3220 & 1.1000 \\
    5228  &-0.1325 & 0.0209 & 0.0363 & 6.2920 &-2.1600 & 4.8060 & 1.7300 & 3.0820 & 0.1529 & 2.5730 & 2.0500 \\
    5248  & 2.4470 &-0.0860 &-0.0652 & 3.5480 & 1.7650 & 1.3220 & 3.3970 & 1.3930 & 0.0838 & 1.4070 & 1.1990 \\
    5249  & 2.2360 &-0.0822 &-0.0638 & 3.7910 & 1.4580 & 2.0730 & 3.3650 & 1.8860 & 0.0899 & 1.6140 & 1.3370 \\
    5284  & 2.4350 &-0.0896 &-0.0691 & 3.3520 & 1.9440 & 1.5680 & 3.5410 & 1.2500 & 0.0718 & 1.4170 & 1.1730 \\
    5318  &-0.5367 & 0.0583 & 0.0783 & 6.2910 &-2.6440 & 5.7050 & 1.6200 & 3.6720 & 0.2010 & 3.0350 & 2.5910 \\
    5342  & 2.4070 &-0.0882 &-0.0683 & 3.8590 & 1.5780 & 1.7010 & 3.3740 & 1.5190 & 0.0850 & 1.4940 & 1.1640 \\
    5350  & 0.3455 &-0.0233 &-0.0069 & 5.8260 &-1.2980 & 4.5590 & 2.3520 & 3.0070 & 0.1533 & 2.5630 & 2.2000 \\
    5362  & 0.7275 &-0.0467 &-0.0328 & 5.6220 &-0.7988 & 3.3720 & 2.3280 & 2.1620 & 0.1239 & 2.0650 & 1.6160 \\
    5451  & 1.5520 &-0.0627 &-0.0474 & 3.5540 & 0.7303 & 2.4130 & 2.2350 & 1.9260 & 0.1018 & 1.8950 & 1.4650 \\
    5544  & 1.4190 &-0.0590 &-0.0376 & 4.8810 & 0.2851 & 2.5190 & 2.6930 & 2.0140 & 0.1047 & 1.7630 & 1.5410 \\
    5571  & 2.9790 &-0.1046 &-0.0779 & 2.5040 & 2.9570 & 1.0840 & 4.1210 & 0.9612 & 0.0517 & 1.3190 & 1.1960 \\
    5583  & 2.2060 &-0.0825 &-0.0601 & 3.6910 & 1.6270 & 1.7870 & 3.3220 & 1.6920 & 0.0722 & 1.5700 & 1.2890 \\
    5586  & 2.4330 &-0.0931 &-0.0688 & 3.1840 & 2.1150 & 1.9050 & 3.5070 & 1.2550 & 0.0708 & 1.4060 & 1.2950 \\
    5643  & 0.8413 &-0.0290 &-0.0105 & 4.7740 &-0.2366 & 3.3870 & 2.2390 & 2.5120 & 0.1412 & 2.3450 & 2.0180 \\
    5667  & 3.8030 &-0.1166 &-0.0924 & 1.1210 & 3.8120 & 0.6684 & 4.6040 & 1.1990 & 0.0626 & 1.1680 & 1.0500 \\
    5679  & 2.3330 &-0.0872 &-0.0662 & 3.7270 & 1.7080 & 1.8900 & 3.4330 & 1.4100 & 0.0767 & 1.3710 & 1.2590 \\
    5688  & 2.3600 &-0.0883 &-0.0635 & 3.2900 & 1.9500 & 1.7650 & 3.5380 & 1.5120 & 0.0789 & 1.4120 & 1.3480 \\
    5699  & 2.1560 &-0.0762 &-0.0557 & 3.6260 & 1.5890 & 1.8640 & 3.4010 & 1.5540 & 0.0812 & 1.4880 & 1.3660 \\
    5716  & 3.7310 &-0.1166 &-0.0910 & 1.4190 & 3.8170 & 0.6320 & 4.5290 & 1.1250 & 0.0577 & 1.2450 & 1.1110 \\
    5741  & 3.5250 &-0.1098 &-0.0828 & 1.6800 & 3.3810 & 0.4970 & 4.3640 & 1.2770 & 0.0729 & 1.3030 & 1.0070 \\
    5756  & 0.4934 &-0.0192 &-0.0015 & 5.8670 &-1.3880 & 3.9920 & 2.2090 & 2.4950 & 0.1438 & 2.2780 & 2.1380 \\
    5790  & 1.9560 &-0.0751 &-0.0563 & 3.8440 & 1.4240 & 2.0180 & 3.2710 & 1.5170 & 0.0716 & 1.6810 & 1.3090 \\
    5795  & 2.3000 &-0.0880 &-0.0638 & 3.8310 & 1.3350 & 1.8720 & 3.3840 & 1.7530 & 0.0916 & 1.6410 & 1.3410 \\
    5833  & 3.5450 &-0.1085 &-0.0840 & 0.6459 & 3.8750 & 0.5400 & 4.3300 & 1.1480 & 0.0648 & 1.0540 & 1.0460 \\
    5853  & 1.3380 &-0.0647 &-0.0478 & 5.0660 & 0.1488 & 2.6760 & 2.8280 & 1.6980 & 0.0841 & 1.8720 & 1.2700 \\
    5855  &-0.5064 & 0.0567 & 0.0814 & 6.2290 &-2.5850 & 5.4600 & 1.5810 & 3.4320 & 0.1819 & 2.7090 & 2.4010 \\
    5927  &-0.1161 &-0.0139 & 0.0042 & 6.1910 &-1.8380 & 4.0610 & 2.0470 & 3.0560 & 0.1669 & 2.6250 & 2.1500 \\
    5929  & 1.8280 &-0.0715 &-0.0521 & 4.3900 & 1.0160 & 2.2430 & 3.1220 & 1.4430 & 0.0758 & 1.6710 & 1.3540 \\
    5969  & 2.7860 &-0.0898 &-0.0680 & 2.8390 & 2.4270 & 1.2450 & 3.6860 & 1.2620 & 0.0506 & 1.4190 & 1.2740 \\
    5993  & 1.1720 &-0.0494 &-0.0339 & 4.8060 &-0.0446 & 2.7430 & 2.6520 & 2.0150 & 0.1135 & 1.8920 & 1.4980 \\
    5996  & 0.6355 &-0.0346 &-0.0239 & 5.5530 &-0.7603 & 3.6020 & 2.4280 & 2.2500 & 0.1211 & 2.0740 & 1.7290 \\
    5997  &-0.7729 & 0.1205 & 0.1478 & 6.2870 &-2.6590 & 5.7530 & 1.5230 & 3.3980 & 0.1942 & 2.9150 & 2.5600 \\
    6089  & 2.0870 &-0.0813 &-0.0615 & 3.8510 & 1.4750 & 2.0260 & 3.2100 & 1.5350 & 0.0633 & 1.6480 & 1.3260 \\
    6107  & 0.6495 &-0.0394 &-0.0272 & 5.2630 &-0.7022 & 3.2150 & 2.4020 & 2.1970 & 0.1223 & 2.0390 & 1.7040 \\
    6114  &-0.6396 & 0.0471 & 0.0607 & 6.3370 &-2.4530 & 5.5260 & 1.7060 & 3.3890 & 0.1808 & 2.7740 & 2.3590 \\
    6134  & 2.2270 &-0.0872 &-0.0637 & 3.5200 & 1.5980 & 1.8430 & 3.4880 & 1.5910 & 0.0891 & 1.5020 & 1.2290 \\
    6158  & 1.3580 &-0.0649 &-0.0489 & 4.5780 & 0.6104 & 2.6430 & 2.6800 & 1.9630 & 0.0940 & 1.7930 & 1.5690 \\
    6169  &-0.5138 & 0.0632 & 0.0806 & 6.0210 &-2.5500 & 5.3930 & 1.5990 & 3.3860 & 0.1934 & 2.8360 & 2.5970 \\
    6228  & 2.0570 &-0.0855 &-0.0689 & 3.9570 & 1.3130 & 1.9620 & 3.2580 & 1.3770 & 0.0657 & 1.5600 & 1.2440 \\
    6259  &-0.9166 & 0.0757 & 0.1007 & 6.4420 &-2.8500 & 5.6510 & 1.5730 & 4.0570 & 0.2143 & 3.0730 & 2.4050 \\
    6313  & 2.2710 &-0.0868 &-0.0654 & 3.3920 & 1.8730 & 1.7610 & 3.5000 & 1.4410 & 0.0748 & 1.5120 & 1.1330 \\
    6395  & 1.9910 &-0.0806 &-0.0597 & 4.2720 & 1.0910 & 2.0080 & 3.0960 & 1.9860 & 0.0891 & 1.5930 & 1.3650 \\
    6408  & 0.2703 &-0.0144 & 0.0028 & 5.7480 &-1.2760 & 4.2950 & 2.0830 & 2.8160 & 0.1345 & 2.3000 & 1.9840 \\
    6470  &-1.4870 & 0.2358 & 0.2844 & 6.2960 &-3.4310 & 7.7730 & 0.9060 & 3.6800 & 0.2921 & 3.8690 & 3.4910 \\
    6471  &-1.1870 & 0.1540 & 0.2084 & 5.8500 &-3.2070 & 8.6350 & 0.7761 & 4.3940 & 0.4100 & 4.1820 & 4.3690 \\
    6472  &-1.1900 & 0.1644 & 0.1912 & 6.7930 &-2.8270 & 5.3920 & 1.4300 & 2.6690 & 0.1739 & 3.0650 & 2.5640 \\
    6477  & 2.1780 &-0.0835 &-0.0606 & 3.7970 & 1.4510 & 1.8820 & 3.3550 & 1.5290 & 0.0851 & 1.6310 & 1.3730 \\
    6480  & 4.0450 &-0.1177 &-0.0893 & 0.8838 & 4.3800 & 0.8779 & 4.8870 & 0.7455 & 0.0436 & 1.2260 & 1.0270 \\
    6481  & 7.0520 &-0.2051 &-0.1683 &-2.9240 & 7.1090 &-0.6073 & 7.1850 & 0.0866 & 0.0205 & 0.0389 & 0.0213 \\
    6482  &-1.3470 & 0.2099 & 0.2611 & 6.2420 &-3.3350 & 7.7600 & 0.8280 & 3.8330 & 0.3172 & 3.9210 & 3.5610 \\
    6484  & 3.9380 &-0.1190 &-0.0912 & 0.4985 & 4.2160 & 0.3073 & 4.6610 & 0.8662 & 0.0595 & 1.0440 & 0.8099 \\
    6485  &-1.2040 & 0.2296 & 0.2623 & 6.7130 &-2.9230 & 6.0340 & 1.5110 & 2.8570 & 0.1915 & 3.3000 & 2.8240 \\
    6486  &-1.3900 & 0.2418 & 0.2834 & 6.3810 &-3.4920 & 7.4740 & 1.0910 & 3.6520 & 0.2707 & 3.7370 & 3.3920 \\
    6488  &-0.3128 & 0.0549 & 0.0778 & 5.8770 &-1.7400 & 4.6280 & 1.8610 & 2.7210 & 0.1600 & 2.5440 & 2.2240 \\
    6489  &-1.2280 & 0.1916 & 0.2265 & 6.5360 &-2.8870 & 6.0840 & 1.3880 & 3.3180 & 0.1968 & 3.2150 & 2.6390 \\
    6490  & 9.0460 &-0.3158 &-0.2714 &-3.7640 & 9.2760 &-1.8830 & 9.0600 & 0.4973 & 0.0164 & 0.2899 & 0.0941 \\
    6491  & 1.5270 &-0.0628 &-0.0464 & 4.1230 & 1.0880 & 2.3480 & 3.0050 & 1.2600 & 0.0634 & 1.6230 & 1.2550 \\
    6492  &-1.3510 & 0.2239 & 0.2554 & 6.7100 &-2.9540 & 6.2510 & 1.5190 & 2.8350 & 0.1745 & 3.2140 & 2.5730 \\
    6494  &-1.0970 & 0.2244 & 0.2599 & 6.6090 &-2.8540 & 5.7160 & 1.5210 & 2.7620 & 0.1688 & 3.1190 & 2.6600 \\
    6495  &-1.1850 & 0.1990 & 0.2486 & 6.1260 &-3.4330 & 8.3990 & 0.6640 & 3.8390 & 0.3584 & 4.1590 & 4.1790 \\
    6497  &-1.3540 & 0.2750 & 0.3102 & 6.0460 &-2.6810 & 6.5930 & 1.4390 & 3.3610 & 0.1985 & 3.4100 & 2.9380 \\
    6499  &-1.3530 & 0.2355 & 0.2879 & 6.4090 &-3.3460 & 7.8350 & 0.9870 & 3.5970 & 0.2929 & 3.8970 & 3.4580 \\
    6501  & 7.3830 &-0.2491 &-0.2039 &-2.1570 & 8.0400 &-1.3630 & 7.9410 & 0.4634 & 0.0290 & 0.8123 & 0.5647 \\
    6502  &-1.2750 & 0.1865 & 0.2135 & 6.6060 &-3.0200 & 6.3420 & 1.4180 & 3.6170 & 0.2331 & 3.2620 & 2.7790 \\
    6503  &-1.4410 & 0.2374 & 0.2736 & 6.7290 &-2.9620 & 6.1870 & 1.4620 & 2.9450 & 0.1806 & 3.2220 & 2.6680 \\
    6505  &-1.1990 & 0.1870 & 0.2184 & 6.3550 &-2.8850 & 6.4050 & 1.4870 & 3.2760 & 0.2101 & 3.3190 & 2.7220 \\
    6506  &-1.3100 & 0.2188 & 0.2536 & 6.5560 &-2.8660 & 5.9900 & 1.5470 & 2.8710 & 0.1701 & 3.1120 & 2.5860 \\
    6510  & 8.9160 &-0.3072 &-0.2640 &-3.4490 & 9.0830 &-1.8710 & 8.8540 & 0.6110 & 0.0198 & 0.3626 & 0.2005 \\
    6512  &-1.3750 & 0.2205 & 0.2540 & 6.5910 &-2.9480 & 5.7710 & 1.5090 & 2.7030 & 0.1632 & 3.2150 & 2.6940 \\
    6513  &-0.9952 & 0.1809 & 0.2239 & 6.4960 &-3.0620 & 6.3080 & 1.0570 & 3.2110 & 0.2281 & 3.3220 & 2.8270 \\
    6514  &-0.9137 & 0.0462 & 0.0918 & 5.0840 &-2.8110 & 7.7780 & 1.3350 & 6.1520 & 0.4777 & 3.9160 & 3.9860 \\
    6515  &-1.4070 & 0.2461 & 0.2927 & 6.3500 &-3.4810 & 7.5920 & 1.0580 & 3.8170 & 0.2846 & 3.8250 & 3.4950 \\
    6516  &-1.3610 & 0.2350 & 0.2659 & 6.6360 &-3.0710 & 5.9680 & 1.5370 & 2.8610 & 0.1869 & 3.2990 & 2.7320 \\
    Errors & 0.03 & 0.007 & 0.0004 & 0.005 & 0.02 & 0.04 & 0.04 & 0.08 & 0.005 & 0.02 & 
0.04 \\
\enddata

\tablecomments{The Id numbers of M67 stars are from Montgomery et
al. (1993).  The last row lists mean errors estimated as described in
Section \ref{conver}}

\end{deluxetable}


\clearpage










\end{document}